\documentclass[12pt,final]{article}
\usepackage{graphicx}
\usepackage{epsfig}
\usepackage{dcolumn}
\usepackage{bm}
\usepackage{achemso}
\usepackage{amssymb,amsmath}
\usepackage{gensymb}
\usepackage{multirow}
\usepackage{booktabs}
\makeatletter
\def\@biblabel#1{(#1)}
\makeatother
\usepackage{lineno}
\usepackage{setspace}

\usepackage{booktabs,rotating}
\usepackage{fancyhdr}
\usepackage{booktabs,caption}
\usepackage[flushleft]{threeparttable}
\usepackage{enumerate,subfigure,tabularx,longtable}
\usepackage{longtable}
\usepackage[utf8x]{inputenc}
\UseRawInputEncoding
\usepackage{gensymb}
\usepackage {threeparttable} 

\usepackage{natbib}
\setkeys{acs}{articletitle = true}

\newcommand{\comment}[1]{}

\def\gsim {\mbox{\hbox{ \lower-.6ex\hbox{$>$}
\kern-1.12em \lower.5ex\hbox{$\sim$}\kern+.35em}}}
\def\lsim {\mbox{\hbox{ \lower-.6ex\hbox{$<$}
\kern-1.12em \lower.5ex\hbox{$\sim$}\kern+.35em}}}
\makeatletter
\let\@fnsymbol\@arabic
\makeatother

\begin{document}


\title{\vspace {-2cm} Estimating Fluid-solid Interfacial Free Energies
	for Wettabilities: A Review of Molecular Simulation Methods}

	\author{Yafan Yang$^{\S,\ddag,1,*}$, Arun Kumar Narayanan Nair$^\|$, \\
		Shuyu Sun$^{\|,1}$, and Denvid Lau$^{\ddag,}$\thanks{Corresponding Authors, email: yafan.yang@cumt.edu.cn; shuyu.sun@kaust.edu.sa; denvid.lau@cityu.edu.hk. \\
		$^*$The original manuscript was submitted to Advances in Colloid and Interface Science in November 2023 and remains under review.} \\
		$^{\S}$State Key Laboratory for Geomechanics \\
		and Deep Underground Engineering, \\
		China University of Mining and Technology, \\
		Xuzhou, Jiangsu, China. \\
		$^\ddag$Department of Architecture and Civil Engineering, \\
		City University of Hong Kong, Hong Kong, China. \\
		$^\|$Computational Transport Phenomena Laboratory,\\
		Physical Science and Engineering Division,\\
		King Abdullah University of Science and Technology, \\
		Thuwal, Saudi Arabia. \\
	}
	\date{\today}
	\maketitle
	\newpage
	\begin{abstract}
		
		Fluid-solid interfacial free energy (IFE) is a fundamental parameter influencing wetting behaviors, which play a crucial role across a broad range of industrial applications.
		Obtaining reliable data for fluid-solid IFE remains challenging with experimental and semi-empirical methods, and the applicability of first-principle theoretical methods is constrained by a lack of accessible computational tools.
		In recent years, a variety of molecular simulation methods have been developed for determining the fluid-solid IFE. 
		This review provides a comprehensive summary and critical evaluation of these techniques.
		The developments, fundamental principles, and implementations of various simulation methods are presented from mechanical routes, such as the contact angle approach, the technique using Bakker's equation, and the Wilhelmy simulation method, as well as thermodynamic routes, including the cleaving wall method, the Frenkel-Ladd technique, and the test-volume/area methods.
		These approaches can be applied to compute various fluid-solid interfacial properties, including IFE, relative IFE, surface stress, and superficial tension, although these properties are often used without differentiation in the literature.
		Additionally, selected applications of these methods are reviewed to provide insight into the behavior of fluid-solid interfacial energies in diverse systems.
		We also illustrate two interpretations of the fluid-solid IFE based on the theory of Navascu{\'e}s and Berry and Bakker's equation. 
		It is shown that the simulation methods developed from these two interpretations are identical. 
		This review advocates for the broader adoption of molecular simulation methods in estimating fluid-solid IFE, which is essential for advancing our understanding of wetting behaviors in various chemical systems.
		
	\end{abstract}
	

	\newpage
	\section{Introduction}
	
	Accurately characterizing wetting interactions at liquid-solid interfaces is essential for optimizing processes in industries such as oil recovery, food production, pharmaceuticals, and coatings, where interfacial phenomena directly influence material performance and process efficiency.\cite{de1985wetting,drelich2019contact}
	Historically, determining the interfacial properties of materials has advanced in parallel with the development of thermodynamic principles, 
	with foundational work by Gibbs\cite{gibbs1906scientific} setting the stage for modern investigations.
	In particular, interfacial free energy (IFE) plays a key role in interface science.\cite{israelachvili2022surface,jian2018atomistic,tam2017moisture,yu2015molecular}
	Understanding IFE is essential for the theoretical explanation of the interfacial phenomenon and for accurate predictions of material behavior in different industrial and engineering applications.
	
	In the three-phase contact region of fluid-fluid-solid systems, IFEs are balanced as described by Young's equation:\cite{young1805iii}
	\begin{equation}
		\gamma_{\rm F_1F_2} \cdot \rm cos\ \theta = \gamma_{\rm SF_2}-\gamma_{\rm SF_1},
		\label{eq:young}
	\end{equation}
	where $\theta$ is the wettability of fluid phase 1 ($\rm F_1$) on solid phase ($\rm S$) surrounded by fluid phase 2 ($\rm F_2$), $\gamma_{\rm F_1F_2}$, $\gamma_{\rm SF_2}$, and $\gamma_{\rm SF_1}$ are the IFEs between phases denoted in the subscript. 
	
	In experiments, although fluid-fluid IFE $\gamma_{\rm F_1F_2}$ and wettability $\theta$ can be directly obtained,\cite{drelich2002measurement,lander1993systematic} the measurement of fluid-solid IFE $\gamma_{\rm SF}$ is difficult and usually inaccurate.\cite{zhu2023experimental}
	Many techniques\cite{geguzin1962surface,linford1972surface,brown1976structure} exist for ascertaining the vacuum-solid IFE and a substantial body of experimental findings has been systematically correlated by Kumikov and Khokonov.\cite{kumikov1983measurement} 
	Meanwhile, due to the limitations of experimental techniques when approaching the nanoscale interfaces, indirect methods have been developed for the fluid-solid IFE of various systems, including cleavage test,\cite{bailey1967direct} solubility test,\cite{parks1984surface} adhesion force measurement,\cite{claesson1986interactions} deformation analysis on the solid film,\cite{nadermann2013solid} contact angle/contact line curvature measurement,\cite{ward2008effect} and contact angle experiment combined with Makkonen hypothesis.\cite{sarkar2023new} 
	A recent review comprehensively discusses both indirect and direct techniques for determining the IFE of fluid-solid interfaces in experiments, highlighting advancements and ongoing challenges.\cite{zhu2023experimental}
	However, precisely determining fluid-solid IFE experimentally remains challenging due to factors such as the size effects, contaminations, and irregularities of surfaces.\cite{tyson1977surface,di2022cleaving,zhu2023experimental}
	
	Semi-empirical theories have been extensively employed to estimate fluid-solid IFE using measured contact angle data.
	These theories encompass approaches like Neumann's equation of state approach,\cite{neumann1974equation,li1992equation} Zisman method,\cite{fox1952spreading} Fowkes method,\cite{fowkes1964attractive} geometric-mean approach,\cite{owens1969estimation} harmonic-mean approach,\cite{wu1971calculation} and van Oss-Good method.\cite{van1986role}
	A comprehensive summary of these semi-empirical theories, along with their assumptions, is provided by \.{Z}enkiewicz.\cite{zenkiewicz2007methods} 
	However, limited research has focused on providing a microscopic foundation for these semi-empirical theories, and significant controversies remain over the validity of several  \textit{ad hoc} assumptions inherent in their formulations.\cite{sullivan1981surface,good1975spreading,yang2023molecular,zhu2023experimental}
	
	First-principle theories have also been developed for determining the interfacial energies. 
	In the statistical mechanical theory of Navascu{\'e}s and Berry\cite{navascues1977statistical}, the Kirkwood-Buff method\cite{kirkwood1949statistical} was extended to the case where fluids are in contact with a rigid solid phase, and the fluid-solid IFE was split into one solid and two fluid-solid contributions. Note that this theory was later combined with molecular simulation\cite{nijmeijer1990wetting} for estimating the fluid-solid relative IFE, which will be discussed in detail in the next section.
	The square gradient theory, which was first introduced by Rayleigh\cite{rayleigh1892xx} and van der Waals\cite{van2012lehrbuch} and later rediscovered by Cahn and Hilliard,\cite{cahn1958free} was applied to study wetting problems.\cite{cahn1977critical,benner1981structure} 
	In square gradient theory, the fluid-solid relative IFE is derived from the surface excess transverse stress, where principal stress profiles are computed based on density distribution across the interface.\cite{benner1981structure,davis1982stress} 
	Meanwhile, classical density functional theory (cDFT) has become a pivotal tool for analyzing wettability phenomena at the molecular level.\cite{ebner1977new,sullivan1981surface,evans1987phase} 
	Within the cDFT framework, 
	the fluid-solid relative IFE is typically calculated using excess grand potential: $ \gamma_{\rm SF}^* = (\Omega + pV)/A$, where $\Omega$, $p$, $V$, and $A$ are the grand potential, bulk pressure, volume of the fluid, and interfacial area, respectively. 
	The cDFT is a robust and versatile tool grounded in thermodynamic principles, 
	offering exceptional accuracy and efficiency.\cite{evans1979nature,wu2007density}
	The cDFT has been used for computing the fluid-solid IFE for a broad range of systems including confined fluid,\cite{yu2009novel} chemically patterned wall,\cite{yatsyshin2018microscopic} and heterogeneous surface,\cite{dkabrowska2022contact} while the application of other first-principle theories has been relatively rare nowadays.
	Despite the demonstrated precision and efficacy of cDFT, it remains underutilized across experimental, theoretical, and computational communities. Major obstacles to its wider adoption include the theoretical complexity and a lack of accessible, user-friendly software for fluid-solid IFE computations.\cite{salinger2008tramonto,jiang2022software,rehner2023feos}
	
	Molecular simulation is a powerful tool for investigating fluid-solid IFE due to several distinct advantages. Firstly, there is a wealth of open-source software available, offering a wide array of simulation tools to choose from.\cite{thompson2022lammps,abraham2015gromacs,hens2020brick} 
	Moreover, molecular simulations are based on precise atomic-level representations, offering a robust physical foundation for IFE calculations.\cite{frenkel2023understanding,allen2017computer}
	Furthermore, molecular simulations provide flexibility, enabling the study of diverse fluid-solid interfaces across a wide range of chemical systems. This versatility arises from the extensive library of force field parameters available in the literature.\cite{heinz2013thermodynamically,cygan2004molecular,schmitt2023extension,senftle2016reaxff,hao2020effect,hao2020carbon} 
	Although certain acceleration techniques, such as coarse-grained modeling,\cite{avendano2011saft,nie2023mesoscale,qin2021role,yu2016mesoscopic,yu2015development} can reduce computational time, molecular simulation methods remain more computationally intensive than semi-empirical and first-principle approaches.
	
	Despite the advancement of various molecular simulation methods for estimating fluid-solid IFE, there remains a scarcity of comprehensive reviews.
	Jiang and Patel\cite{jiang2019recent} published a review paper focusing on molecular simulation techniques for estimating contact angles. Those methods are useful for calculating the differences of IFE. 
	Nevertheless, several essential methodologies for the direct estimation of fluid-solid IFE remain uncovered in the previous review, and numerous innovative techniques have emerged over the past few years.
	This review aims to bridge these gaps by providing a comprehensive analysis of state-of-the-art molecular simulation methodologies for calculating fluid-solid IFE.
	The molecular simulation methods for estimating IFE can be broadly classified into two primary categories: mechanical and thermodynamic approaches.
	The mechanical approach encompasses methods such as the contact angle approach, the method using Bakker's equation, and the Wilhelmy simulation method, each of which will be discussed in detail. For the thermodynamic approach, key methods include the cleaving wall technique, the Frenkel-Ladd technique, and the test-volume/area methods.
	
	Note that while there are various experimental\cite{bokeloh2011nucleation,aman2016interfacial,granasy2002interfacial,hu2011single,louhichi2013nucleation,djikaev2017self} and theoretical\cite{hartel2012tension,wang2013density,schoonen2022crystal} methods available for studying fluid-solid IFE in the context of crystallization/nucleation, the focus of this review is on wetting problems and methods that are not suitable for studying common wetting problems were not included in the above discussion. 
	Therefore, this review will not cover molecular simulation techniques such as the capillary fluctuation technique,\cite{hoyt2001method}
	umbrella sampling,\cite{auer2001prediction}
	extrapolation method,\cite{ferreira2000temperature,barroso2002solid} 
	metadynamics,\cite{laio2002escaping,angioletti2010solid,lau2014robust,lau2012characterization}
	the superheating and undercooling method,\cite{luo2003maximum,luo2004nonequilibrium,luo2005deducing}
	the seeding technique,\cite{bai2005differences,knott2012homogeneous,sanz2013homogeneous}
	tethered Monte Carlo,\cite{fernandez2012equilibrium}
	and the mold integration method\cite{espinosa2014mold}.
	
	\section{Mechanical Routes}
	The mechanical route is based on the direct computation of forces at the interface within molecular simulations. This route includes the contact angle approach,\cite{maruyama2002molecular} the technique based on Bakker's equation,\cite{nijmeijer1990wetting} and the Wilhelmy simulation method.\cite{imaizumi2020wilhelmy} 
	
	\subsection{Contact Angle Approach}
	While the contact angle approach does not provide direct access to the absolute values of fluid-solid IFE $\gamma_{\rm SF}$, the method is still noteworthy considering its capability to estimate the differences of fluid-solid IFE ($i.e.,$ $\gamma_{\rm SF_2}-\gamma_{\rm SF_1}$ in Eq. \ref{eq:young}). 
	This property, termed ``superficial tension" by Gibbs\cite{toshev2006wetting,valiya2022comment} (sometimes also noted as ``adhesion tension"\cite{defay1966surface}), describes the difference in fluid-solid IFE across different phases.
	Notably, when the fluid phase 2 is vapor with low density and the interfacial adsorption of the vapor is little, $\gamma_{\rm SF_2}$ can be used to approximate the vacuum-solid IFE. 
	In this case, the relative IFE between fluid phase 2 and the solid ($i.e.,$ $\gamma_{\rm SF_2}^*$ defined in the next section) is close to zero, and the relative IFE between fluid phase 1 and the solid can be estimated as the negative of the superficial tension.\cite{taherian2013contact,fan2020microscopic} 
	The method involves calculation of fluid-fluid IFE $\gamma_{\rm F_1F_2}$ and contact angle $\theta$ in separate simulations, and superficial tension is calculated as the product of $\gamma_{\rm F_1F_2}$ and $\rm cos\ \theta$, based on Eq. \ref{eq:young}.\cite{yang2022interfacial,yang2023molecular} 
	
\begin{figure}[tb]
		\begin{centering}
			\includegraphics[width=0.65\textwidth]{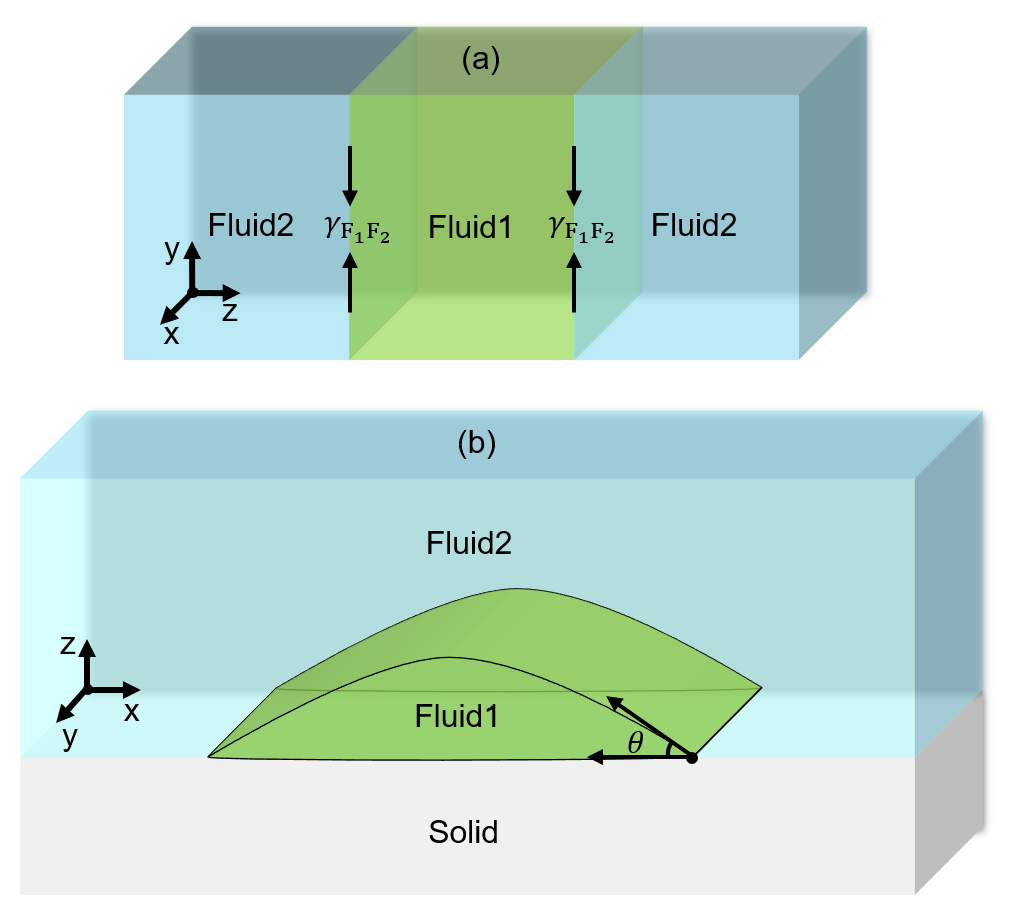}
			\caption{System setups for estimating (a) the fluid-fluid IFE and (b) the wettability in the contact angle approach.}
			\label{fig:z_f1}
		\end{centering}
	\end{figure}

	%
	The sketches of molecular system setups for estimating $\gamma_{\rm F_1F_2}$ and $\theta$ are shown in Fig. \ref{fig:z_f1}a and b, respectively.
	To calculate fluid-fluid IFE in molecular simulation, the simulation box usually contains slabs of bulk liquid phases to form interfaces (see Fig. \ref{fig:z_f1}a).\cite{ghoufi2016computer,muller2020guide} 
	In an orthogonal system with z-direction normal to the flat fluid interface, the principle components of the pressure tensor ($P_{\rm xx}$, $P_{\rm yy}$, and $P_{\rm zz}$) are used to calculate IFE according to Bakker's equation:\cite{green1960molecular,bakker1928kapillaritat}
	\begin{equation}
		\gamma_{\rm F_1F_2} = \int \Big [P_{\rm zz}-\frac{1}{2}(P_{\rm xx}+P_{\rm yy})\Big ] dz,
		\label{eq:bakker}
	\end{equation}

	To estimate the contact angle, the molecular system containing a sessile droplet on a solid substrate is simulated, and the contact angle can be determined from the average shape of the droplet (see Fig. \ref{fig:z_f1}b).\cite{maruyama2002molecular} 
	It is important to note that the direct simulation of such a three-phase system may be subject to errors including finite size effects\cite{kanduvc2017going,silvestri2019wetting} and hysteresis.\cite{godawat2008structure,macdowell2006nucleation} To circumvent these issues, a large cylindrical shaped droplet is suggested\cite{silvestri2019wetting} and enhanced sampling method\cite{jiang2019characterizing} can be combined.
	Jiang and Patel\cite{jiang2019recent} summarized those issues and provided a comprehensive review on methods for contact angle calculation in molecular simulations.
	Additionally, the contact angle values predicted using molecular simulations are dependent on the choice of the force field. 
	For example, using two different force fields, the water contact angle on calcite is predicted to be 0\degree \ and about 38\degree \ at 323 K and 20 MPa.\cite{silvestri2019wetting,le2021effects} 
	It also is worth mentioning that a freely available tool named ``ContactAngleCalculator" has been recently developed to rapidly and accurately estimate contact angles from molecular simulations.\cite{wang2022contactanglecalculator}

	The molecular simulation data on $\gamma_{\rm F_1F_2}$ or $\theta$ of different systems is extensive.\cite{ghoufi2016computer,stephan2019thermophysical,pan2020review,shi2009molecular,tenney2014molecular,chen2015water} Although superficial tension can be estimated based on those data, direct analysis of its behaviors remains relatively scarce. Few authors have investigated superficial tensions in water/rock, water/gas/rock, water/oil/rock, and water/gas/oil/rock systems. \cite{yang2022interfacialCO2,yang2023molecularH2,yao2024interfacial,yang2022interfacial,cui2023interfacial,yang2022interfacial,yang2022interfacialaromtaic,yang2024interfacial} Such systems are important for applications including gas storage and enhanced gas/oil recovery.\cite{pan2021underground,arif2019wettability,wang2023multiphysical,chen2023visualizing} Remarkably, superficial tension appears as the denominator of the capillary number $N_{\rm Ca}$:
	\begin{equation}
		N_{\rm Ca} = \eta v/(\gamma_{\rm F_1F_2} \cdot {\rm cos}\ \theta),
	\end{equation}
	where $\eta$ and $v$ are the viscosity and characteristic velocity, respectively.
	The capillary number, representing the ratio of viscous to capillary force, is crucial in enhanced gas/oil recovery.\cite{chatzis1984correlation,guo2020critical}
	Note that superficial tension also appears in the Young-Laplace equation for the capillary pressure. Many authors have reported the values of capillary pressure from molecular simulations.\cite{amarasinghe2014molecular,chen2017wettability,chen2019model,sun2020molecular,yang2022interfacialCO2} 
	
\begin{figure}[tb]
		\begin{centering}
			\includegraphics[width=0.6\textwidth]{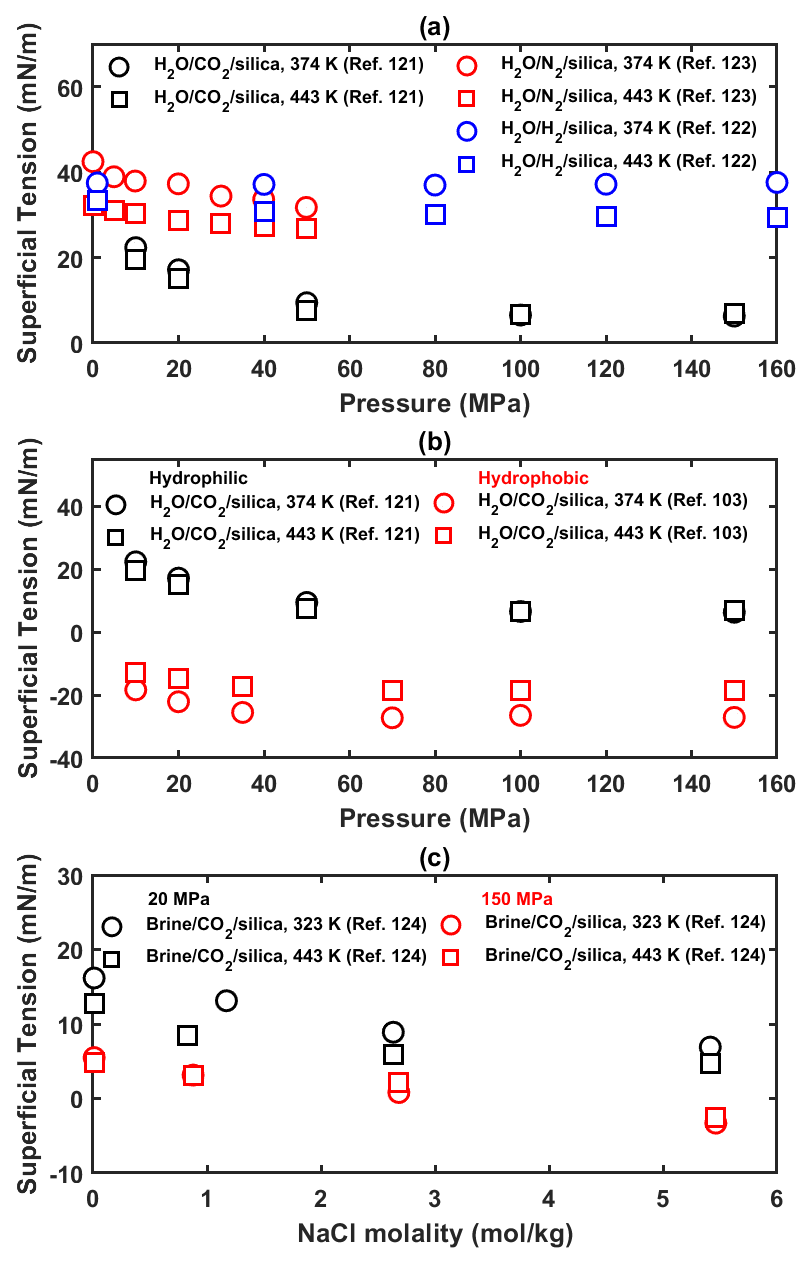}
			\caption{Dependence of superficial tensions on pressure at different temperatures in (a) H$_2$O/gas(CO$_2$, N$_2$, and H$_2$)/hydrophilic silica systems and (b) H$_2$O/CO$_2$/silica (with hydrophilic and hydrophobic surfaces) systems. 
				(c) Dependence of superficial tensions on salinity at different temperatures and pressures in the brine/CO$_2$/hydrophilic silica systems. 
				The data are taken from Refs.\citenum{yang2022interfacial}, \citenum{yang2022interfacialCO2},\citenum{yang2023molecularH2}, \citenum{yao2024interfacial}, and \citenum{cui2023interfacial}.}
			\label{fig:z_f2}
		\end{centering}
	\end{figure}

	Here, we provide a summary of the reported superficial tensions in water/gas/silica systems,\cite{yang2022interfacialCO2,yao2024interfacial,yang2023molecularH2,cui2023interfacial,yang2022interfacial} as depicted in Fig. \ref{fig:z_f2}.
	This serves as an illustrative example of the behavior of superficial tension under varying conditions such as temperature, pressure, gas type, surface wettability, and salinity.
	In general, an increase in pressure results in a decrease in superficial tension, while higher temperatures weaken the magnitude of superficial tension. 
	Notably, the reduction of superficial tension induced by high pressure is significantly more pronounced in systems with CO$_2$ compared to those with N$_2$ or H$_2$ (see Fig. \ref{fig:z_f2}a). 
	This phenomenon can be ascribed to the strong interactions of CO$_2$-H$_2$O and CO$_2$-silica pairs. Interfacial density distribution results indicate that CO$_2$ exhibits greater adsorption in the gas-H$_2$O and gas-silica interfaces compared to N$_2$ or H$_2$. This results in a more pronounced reduction of $\gamma_{\rm F_1F_2}$ and ${\rm cos}\ \theta$.\cite{yang2022interfacialCO2,yang2023molecularH2,yao2024interfacial}
	Adjusting the density of surface silanol groups can modify silica's wettability from hydrophilic to hydrophobic.\cite{emami2014force,chen2015water}
	Experimental modifications to the silica surface, such as the ionization of surface silanol groups, are achievable, particularly through controlled temperature adjustments.\cite{lamb1982controlled}
	Superficial tensions are positive in systems with hydrophilic silica, while negative superficial tensions have been reported in systems with hydrophobic silica surfaces due to the change in the sign of ${\rm cos}\ \theta$ (see Fig. \ref{fig:z_f2}b).
	Additionally, an increase in salinity leads to a reduction in superficial tension (see Fig. \ref{fig:z_f2}c). This reduction comes mainly from the decrease in ${\rm cos}\ \theta$, given that $\gamma_{\rm F_1F_2}$ increases with salinity.

	\subsection{Technique using Bakker's Equation}
	
	While Bakker's original equation (Eq. \ref{eq:bakker}) was designed specifically for fluid-fluid interfaces, it has been adapted from fluid-fluid to fluid-solid interfaces to quantify surface stress $s$ in systems with flexible solid substrates using the same methods for the fluid-fluid interface (see section 2.1).\cite{baidakov2017surface,di2020shuttleworth,wu2021calculation,dreher2019anisotropic,velazquez2023effective}
	Note that the calculated $s$ in this direct approach can be separated into two terms according to the Shuttleworth equation:\cite{shuttleworth1950surface} 
	\begin{equation}
		\label{eq:shuttle}	
		s = \gamma_{\rm SF_1} + A \cdot \partial \gamma_{\rm SF_1}/\partial A.
	\end{equation}
	Here, the fluid-solid IFE is represented by the first term, while the second term quantifies strain-related energy contributions from the solid substrate.\cite{makkonen2012misinterpretation} 
	The second term separates the IFE of solid interfaces from that of liquid interfaces and can be estimated by a numerical derivative using IFE data under slightly varied strains.\cite{di2020shuttleworth} 
	Note that ``surface tension" could be used synonymically with ``surface stress", although a subtle difference exists.\cite{di2020shuttleworth}
	%
	
	In the rest of this section, we present the technique using Bakker's equation for estimating the fluid-solid relative IFE:\cite{yamaguchi2019interpretation}
	\begin{equation}
		\label{eq:RSF}	
		\gamma_{\rm SF_1}^* = \gamma_{\rm SF_1}-\gamma_{\rm S},
	\end{equation}
	where $\gamma_{\rm S}$ is the IFE of the solid substrate in the absence of any fluid.
	Note that both superficial tension and relative IFE represent differences in interfacial energies. 
	In this work, relative IFE is considered as a specific case of negative superficial tension. This distinction is based on the reference state: relative IFE is defined with respect to the vacuum-solid interface, whereas superficial tension may be defined with either vacuum-solid or fluid-solid states.
	%
	The technique using Bakker's equation usually deals with molecular systems containing rigid solids or solids that can be treated as external potential. 
	Flexible solids can in principle be handled. 
	Note that $\gamma_{\rm S}$ cannot be determined within this method. 
	While $\gamma_{\rm S}$ is not directly computable in this method, it can be estimated using complementary techniques including the Frenkel-Ladd techniques,\cite{frenkel1984new,polson2000finite,vega2007revisiting,reddy2018calculation} the cleaving wall methods,\cite{broughton1983surface,broughton1986molecular,tipeev2021direct,di2022cleaving} $\gamma$-integration techniques,\cite{smith1999determining,grochola2002simulation,modak2016determination} and approach based on the generalized Gibbs adsorption equation.\cite{baidakov2017surface}
	The fluid-solid relative IFE can be used to calculate wettability and adhesion work since $\gamma_{\rm S}$ could be canceled out when subtracting $\gamma_{\rm SF_1}$ from $\gamma_{\rm SF_2}$.\cite{fan2021generalized,yang2023molecular}

\begin{figure}[tb]
		\begin{centering}
			\includegraphics[width=0.5\textwidth]{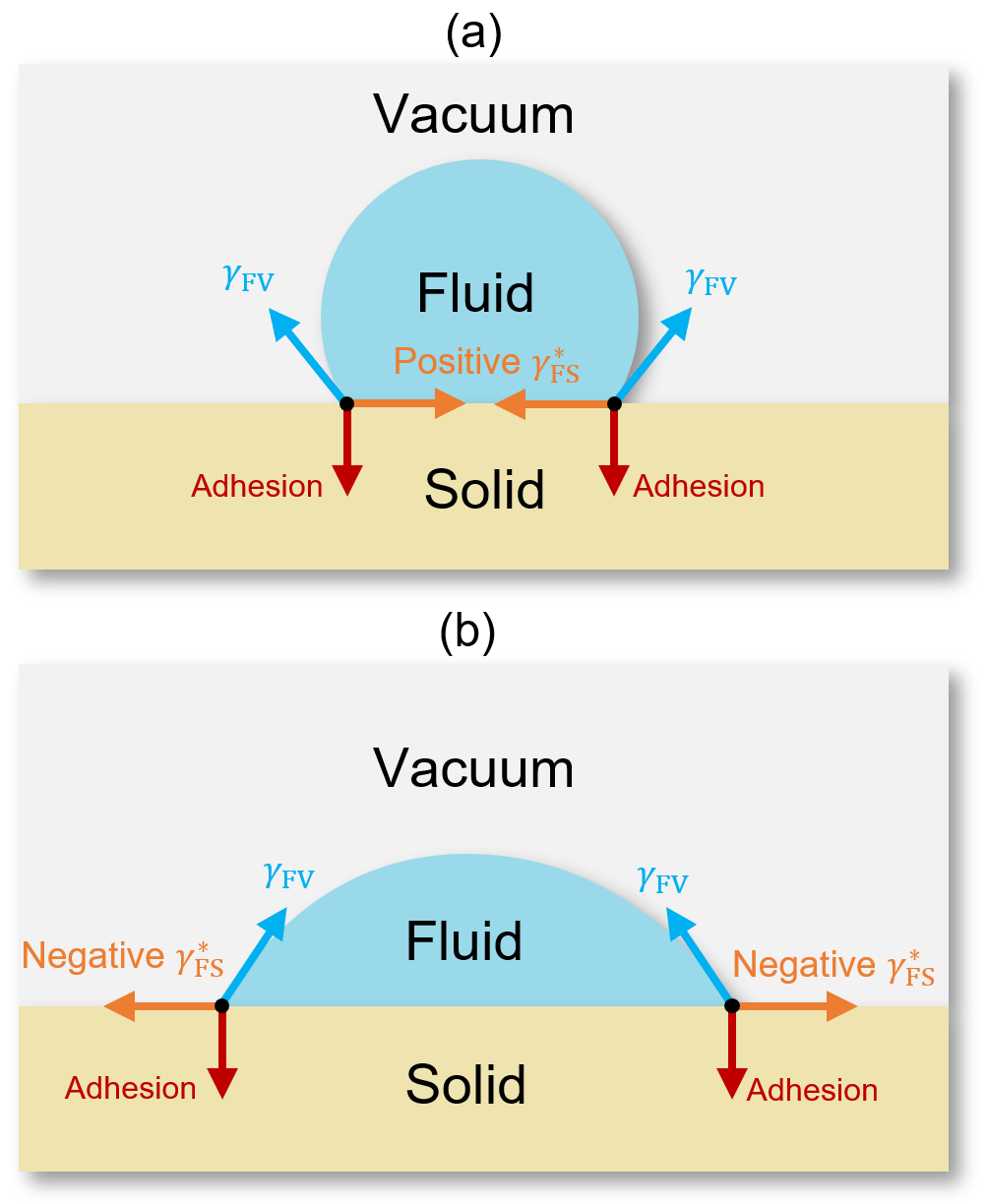}
			\caption{Mechanical balances in phase contact regions in the vacuum/fluid/solid systems with (a) positive fluid-solid relative IFE and (b) negative fluid-solid relative IFE.}
			\label{fig:z_f3}
		\end{centering}
	\end{figure}

	Note that for an interface in equilibrium, the value of IFE $\gamma$ is always positive according to the principles of thermodynamic stability.\cite{firoozabadi2016thermodynamics,ruckenstein1978origin,okazawa1979thermodynamically}
	However, negative values of relative IFE $\gamma_{\rm SF}^*$ are frequently reported.\cite{surblys2014molecular,yang2023molecular}
	Negative $\gamma_{\rm SF}^*$ does not violate the thermodynamic stability as the $\gamma_{\rm S}$ is not included. 
	The sign of $\gamma_{\rm SF}^*$ indicates the direction of tension. Positive $\gamma_{\rm SF}^*$ arises from contracting tension to reduce the contact area (similar to fluid-fluid interfaces), while negative $\gamma_{\rm SF}^*$ comes from the spreading tension to increase the contact area. 
	This can be understood from mechanical balance across phase boundaries in the vacuum/fluid/solid systems by assuming the existence of a vacuum/solid interface ($i.e.,$ the relative IFE of the vacuum/solid interface is 0 mN/m). 
	If the contact angle is greater than 90$\degree$, $\gamma_{\rm SF}^*$ has to be contracting force (positive) to balance the fluid-vacuum IFE $\gamma_{\rm FV}$ in the tangential direction of the surface (see Fig. \ref{fig:z_f3}a).
	While spreading force (negative $\gamma_{\rm SF}^*$) is present when the contact angle is less than 90$\degree$ (see Fig. \ref{fig:z_f3}b). 

\begin{figure}[tb]
		\begin{centering}
			\includegraphics[width=1.0\textwidth]{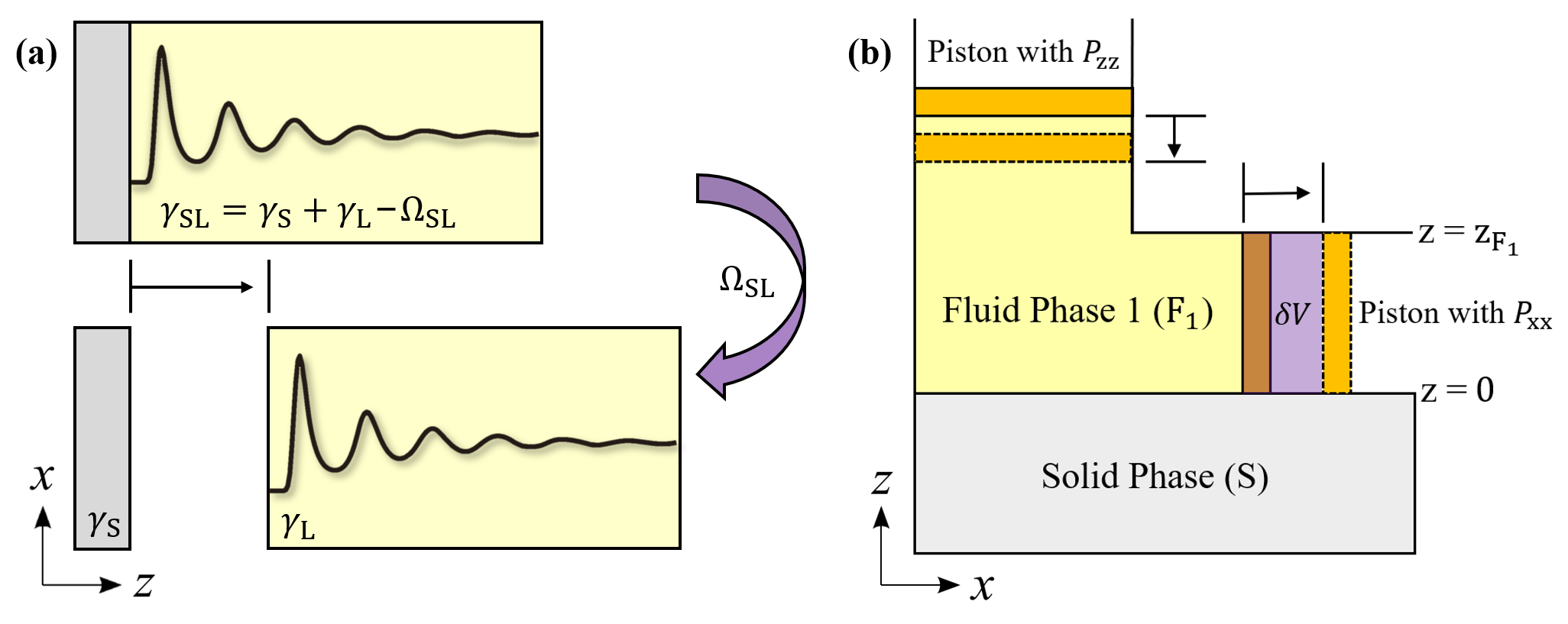}
			\caption{(a) Physical meanings of contributions in fluid-solid IFE from the theory of  Navascu{\'e}s and Berry\cite{navascues1977statistical}. Grey and yellow regions are solid and fluid phases, separately. The solid curve denotes the density distribution of fluid near the solid substrate. Similar figures were presented in Refs.\citenum{navascues1977statistical} and \citenum{fan2020microscopic}. 
				(b) Thought experiment for Bakker's equation applied to the fluid-solid interface. Figure (b) is adapted from Ref.\citenum{yamaguchi2019interpretation}.}
			\label{fig:z_f4}
		\end{centering}
	\end{figure}

	Navascu{\'e}s and Berry\cite{navascues1977statistical} extended the statistical mechanical theory of Kirkwood and Buff\cite{kirkwood1949statistical} to the case where fluids are in contact with a rigid solid phase. 
	Within this theory, the fluid-solid IFE can be understood through a process that combines a vacuum-solid interface and a vacuum-liquid interface into a liquid-solid interface.
	The fluid-solid IFE was split into one solid and two fluid-solid contributions:\cite{navascues1977statistical}
	\begin{equation}
		\label{eq:SF}	
		\gamma_{\rm SF_1} = \gamma_{\rm S} + \gamma_{\rm F_1} -\Omega_{\rm SF_1}.
	\end{equation}
	Here $\gamma_{\rm F_1}$ is the IFE of the fluid phase $\rm F_1$ when the solid phase $\rm S$ is removed without relaxing the structure of $\rm F_1$, which can be calculated using Eq. \ref{eq:bakker}. 
	And $\Omega_{\rm SF_1}$ denotes the free energy change, interpretable as the work required to separate $\rm S$ and $\rm F_1$ without relaxing $\rm F_1$:\cite{navascues1977statistical}  
	\begin{equation}
		\label{eq:OMEGA}
		\Omega_{\rm SF_1} = \int_{0}^{\rm z_{F_1}} {\rm zF_z(z)n(z)} \rm d{\rm z},
	\end{equation}
	where $\rm F_z(z)$ and $\rm n(z)$ are the force exerted	on fluid by solid in the z-direction ($i.e.,$ direction normal to the surface) and number density, respectively. The origin point is defined at the furthest location where the fluid density is zero from the solid surface and $\rm z_{F_1}$ is the point in the bulk region of $\rm F_1$. 
	Then the expression for $\gamma_{\rm SF_1}^*$ is as follows: 
	\begin{equation}
		\gamma_{\rm SF_1}^* = \gamma_{\rm F_1} -\Omega_{\rm SF_1}.
	\end{equation}
	The physical meaning of each term in Eq. \ref{eq:SF} is shown in Fig. \ref{fig:z_f4}a and similar figures were given in Refs.\citenum{navascues1977statistical} and \citenum{fan2020microscopic}. 
	The theory of Navascu{\'e}s and Berry\cite{navascues1977statistical} was later combined with molecular dynamics simulation for estimating $\gamma_{\rm SF_1}^*$.\cite{fan2020microscopic,xu2022effects,yang2023molecular,leroy2015dry}
	
	Nijmeijer and Leeuwen\cite{nijmeijer1990microscopic} also derived an expression for the fluid-solid relative  IFE based on Bakker's equation:
	\begin{equation}
		\gamma_{\rm SF_1}^* = \int_{0}^{\rm z_{F_1}} (P_{\rm zz}-P_{\rm xx}) dz,
		\label{eq:bakker2}
	\end{equation}
	where $P_{\rm xx}$ and $P_{\rm zz}$ are the principal components of the pressure tensor of the fluid considering the solid substrate as external potential.

	This equation can be understood by a thought experiment designed by Yamaguchi et al.\cite{yamaguchi2019interpretation} as shown in Fig. \ref{fig:z_f4}b. This thought experiment is an extension of the thought experiment for Bakker's description of the
	relationship between the IFE and the fluid stress anisotropy.\cite{green1960molecular,bakker1928kapillaritat}
	In this thought experiment, one piston is positioned perpendicular to the fluid-solid interface, covering the area where the fluid is present. This ranges from $z=0$ (where the fluid density is 0 g/cm$^3$) to $z=z_{\rm F_1}$ (a location within the bulk). Another piston is placed parallel to the fluid-solid interface, far from the interface, to regulate the bulk pressure $P_{\rm zz}$.
	By making simultaneous, infinitesimal virtual displacements of the pistons, we can alter only the interface area without affecting the fluid volumes.
	Let's denote the depth normal to the $xz$-plane as $l$. If $\delta V$ represents the infinitesimal volume change caused by the downward displacement of the top piston, and $\delta x$ represents the corresponding displacement of the side piston, then we can deduce the following:\cite{yamaguchi2019interpretation}
	\begin{equation}
		\delta V = l\delta x\int_{0}^{\rm z_{F_1}}dz.
	\end{equation}
	Assuming that the displacement occurs in a quasi-static manner under a constant temperature, the minimum mechanical work, denoted as $\delta W$, needed for this change is linked to the alteration in the Helmholtz energy $F$, as expressed by the following equation:\cite{yamaguchi2019interpretation}
	\begin{equation}
		\delta F = \delta W = P_{\rm zz}\delta V - l\delta x\int_{0}^{\rm z_{F_1}} P_{\rm xx}(z)dz.
	\end{equation}
	The outcome of this change is that the solid-vacuum interface is supplanted by the solid-fluid interface. Consequently, Eq. \ref{eq:bakker2} can be reestablished by taking the partial derivative of the Helmholtz energy $F$ with respect to the fluid-solid area $A_{\rm SF_1}$, where $\delta A_{\rm SF_1} = l\delta x$:\cite{yamaguchi2019interpretation}
	\begin{equation}
		\gamma_{\rm SF_1}-\gamma_{\rm S} = \gamma_{\rm SF_1}^* = \Big(\cfrac{\partial F}{\partial A_{\rm SF_1}}\Big)_{\rm N,V,T} = \int_{0}^{\rm z_{F_1}} (P_{\rm zz}-P_{\rm xx}(z)) dz.
	\end{equation}

	The Eq. \ref{eq:bakker2} was implemented in molecular dynamics simulations for computing $\gamma_{\rm SF_1}^*$.\cite{nijmeijer1990wetting,yamaguchi2019interpretation,yen2020influence} 
	Importantly, the expression of $\gamma_{\rm SF_1}^*$ from the theory of Navascu{\'e}s and Berry\cite{navascues1977statistical} and the derivation of Nijmeijer and Leeuwen\cite{nijmeijer1990microscopic} are identical. 
	The $P_{\rm xx}$ in Eq. \ref{eq:bakker2} is the same as the pressure component tangential to the interface in $\gamma_{\rm F_1}$, and the $P_{\rm zz}$ in Eq. \ref{eq:bakker2} is the summation of the negative of the integrand of $\Omega_{\rm SF_1}$ and the pressure component normal to the interface in $\gamma_{\rm F_1}$.
	However, the agreement was rarely noted.\cite{tang1995fluid,nijmeijer1996comment} For instance, the work of Nijmeijer and Leeuwen\cite{nijmeijer1990microscopic} was overlooked by several authors.\cite{fan2020microscopic,xu2022effects,yang2023molecular,leroy2015dry}
	
	We now outline the computation details required for estimating $\gamma_{\rm SF_1}^*$ in molecular simulations.
	The setup of the simulation box resembles the top figure in Fig. \ref{fig:z_f4}a. An atomistic piston could be implemented to regulate the bulk pressure.\cite{yamaguchi2019interpretation,yang2023molecular} 
	The number density $\rm n(z)$ is readily accessible in simulations and can be calculated on the fly or in post-processing. 
	Once $\rm n(z)$ converges, the locations of the origin point and $\rm z_{F_1}$ can be identified. 
	Without modifying the simulation code, the $\rm F_z(z)$ is calculated by rerunning the saved trajectories with fluid self-interactions disabled.
	With the above properties calculated, $\Omega_{\rm SF_1}$ can be obtained by numerical integration using Eq. \ref{eq:OMEGA}. 
	
	The $\gamma_{\rm F_1}$ can be calculated using Eq. \ref{eq:bakker} by rerunning the saved trajectories with solid substrate deleted from the simulation box. 
	This task usually involves the calculation of the microscopic profile of the pressure tensor.
	The LAMMPS package offers built-in features to determine local pressure tensors.\cite{thompson2022lammps} There are several other analysis tools for this task including the GROMACS-LS code prepared by Vanegas et al.,\cite{torres2016geometric} the MDStressLab tool developed by Admal et al.,\cite{admal2010unified} and the LAMMPS patch developed by Nakamura et al.\cite{nakamura2015precise}
	It is important to note that the definition of microscopic pressure tensor remains under debate, particularly in terms of its contour formalism.\cite{shi2023perspective}
	Different definitions were used in previous studies for estimating fluid-solid IFE.\cite{nishida2014molecular,yamaguchi2019interpretation,yen2020influence,yang2023molecular} 
	However, further investigation is required to determine whether different methods yield consistent $\gamma_{\rm SF_1}^*$.
	Another issue is related to the challenges for systems with many-body and long-range potentials for the computation of local pressure tensors.\cite{nakamura2015precise}
	Additionally, the calculation of pressure tensor is a computationally intensive task given that it involves a second-order property with a slow convergence rate.
	The readers are referred to the perspective paper by Shi et al.\cite{shi2023perspective} for more information regarding the calculation of microscopic pressure tensors using molecular simulations.
	
	Bakker’s equation has been widely utilized to estimate the relative IFE at fluid-solid interfaces in many studies.\cite{nijmeijer1990microscopic,hamada2009phase,nishida2014molecular,fujiwara2014local,dreher2018calculation,yamaguchi2019interpretation,yen2020influence,kaur2021pressure,yang2023molecular}
	For example, Yang et al.\cite{yang2023molecular} applied the method to study the relative IFE in water/gas/solid three-phase systems at various temperatures, pressures, and wettabilities.
	Note that the water wettability on a solid surface depends on the water-solid interaction energy.\cite{theodorakis2015modelling,herdes2018combined} 
	A binary interaction parameter $\rm k_{ij}$ was introduced to adjust the interaction energy between water particle and solid particle: $\varepsilon_{\rm ij} = (1-{\rm k_{ij}})\varepsilon_{\rm ij}^{\rm Mix}$, where $\varepsilon_{\rm ij}^{\rm Mix}$ is the potential well depth obtained from the mixing rule.\cite{lafitte2013accurate}
	
\begin{figure}[tb]
		\begin{centering}
			\includegraphics[width=0.8\textwidth]{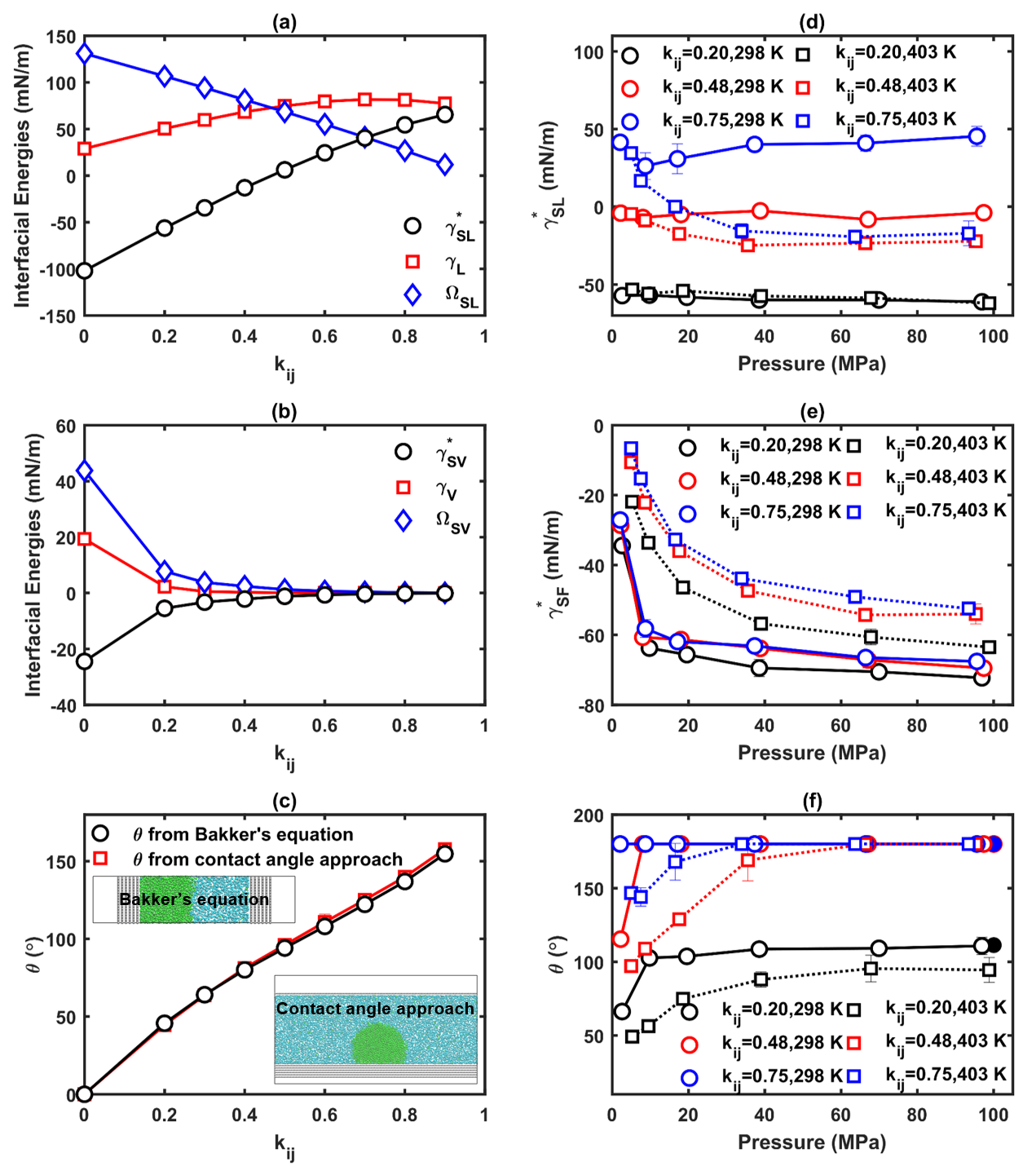}
			\caption{Dependence of interfacial energies for (a) solid-liquid interface, (b) solid-vapor interface, and (c) contact angles $\theta$ (from Bakker's equation and the contact angle approach) on the binary interaction parameter of the water-solid pair k$_{\rm ij}$ in the H$_2$O (vapor/liquid)/solid system at 298 K.
				Dependence of (d) solid-water relative IFE $\gamma_{\rm SL}^*$, (e) solid-CO$_2$ relative IFE $\gamma_{\rm SF}^*$, and (f) $\theta$ on pressure at various k$_{\rm ij}$ and temperatures in the CO$_2$/water/solid system.
				The data are taken from Ref.\citenum{yang2023molecular}.}
			\label{fig:z_f5}
		\end{centering}
	\end{figure}

	The influence of $\rm k_{ij}$ on interfacial energies and wettabilities in the water (vapor/liquid)/solid system is depicted in Fig. \ref{fig:z_f5}a-c.
	It is observed that the relative IFE of the liquid H$_2$O-solid interface $\gamma_{\rm SL}^*$ escalates from -102.0 to 65.0 mN/m, as $\rm k_{ij}$ varies from 0.0 to 0.9. 
	Hydrophilic surfaces correspond to negative $\gamma_{\rm SL}^*$, while hydrophobic surfaces correspond to positive $\gamma_{\rm SL}^*$ (c.f. Fig. \ref{fig:z_f5}a and c).
	Simultaneously, the magnitude of the relative IFE of the vapor H$_2$O-solid interface $\gamma_{\rm SV}^*$ is typically less than that of $\gamma_{\rm SL}^*$.
	$\gamma_{\rm SV}^*$ ascends with $\rm k_{ij}$ and rapidly converges to 0 mN/m.
	Furthermore, wettability can be determined based on $\gamma_{\rm SL}^*$ and $\gamma_{\rm SV}^*$ using Young's equation. The wettabilities derived from Young's equation using relative IFE values from Bakker's equation align well with those obtained from the contact angle approach, as illustrated in Fig. \ref{fig:z_f5}c.
	
	The relative IFEs and wettabilities in the water/CO$_2$/solid system are depicted in Fig. \ref{fig:z_f5}d-f.
	Three distinct values of $\rm k_{ij}$ for the H$_2$O-solid pair are selected to exemplify hydrophilic ($\theta = 45.8\degree,\rm k_{ij} = 0.20$), neutral ($\theta = 91.9\degree,\rm k_{ij} = 0.48$), and hydrophobic ($\theta = 132.7\degree,\rm k_{ij} = 0.75$) surfaces.
	The impact of temperature and pressure on the relative IFE of the water-solid interface $\gamma_{\rm SL}^*$ is complex (see Fig. \ref{fig:z_f5}d).
	Generally, the influence of pressure on $\gamma_{\rm SL}^*$ is less noticeable at 298 K.  However, $\gamma_{\rm SL}^*$ of water+CO$_2$+hydrophobic solid ($i.e.,$, $\rm k_{ij} = 0.75$) system initially decreases and then increases as pressure varies from approximately 5 to 100 MPa. Conversely, at a higher temperature of 403 K, $\gamma_{\rm SL}^*$ diminishes with increasing pressure, and the effect of pressure is more pronounced with more hydrophobic solids.
	In addition, the relative IFE for the CO$_2$-solid interface $\gamma_{\rm SF}^*$ is negative (see Fig. \ref{fig:z_f5}e). The magnitude of $\gamma_{\rm SF}^*$ intensifies as pressure increases and the solid surface becomes more hydrophilic. High temperatures generally reduce the magnitude of $\gamma_{\rm SF}^*$. Furthermore, the contact angles enlarge as pressure rises and temperature decreases for all three surface types (see Fig. \ref{fig:z_f5}e).
	
	These simulation results provide a basis to evaluate the assumptions in semi-empirical theories like the Neumann equation of state, which is commonly used to infer fluid-solid IFEs from fluid-fluid IFE and wettability data.\cite{li1992equation} 
	In H$_2$O/gas/solid systems, the Neumann equation of state is typically combined with the presumption that the H$_2$O-solid IFE remains unchanged across different pressure levels.\cite{ameri2013investigation,abdulelah2021co2,pan2020interfacial,hosseini2022assessment}
	However, simulation computed $\gamma_{\rm SL}^*$ values in water/gas/solid systems suggest that this presumption may not hold true in cases that feature non-hydrophilic surfaces with strong gas-solid interactions due to the adsorption of gases in the water-solid interface.\cite{yang2023molecular}
	
	\subsection{Wilhelmy Simulation Method}
	
	Based on the experimental Wilhelmy method\cite{wilhelmy1863ueber} for measuring the interfacial tension of fluid-fluid interface, 
	Imaizumi et al.\cite{imaizumi2020wilhelmy} developed the Wilhelmy simulation method for estimating interfacial properties in fluid-fluid-solid three-phase systems using a single molecular dynamics simulation. 
	
\begin{figure}[tb]
		\begin{centering}
			\includegraphics[width=0.9\textwidth]{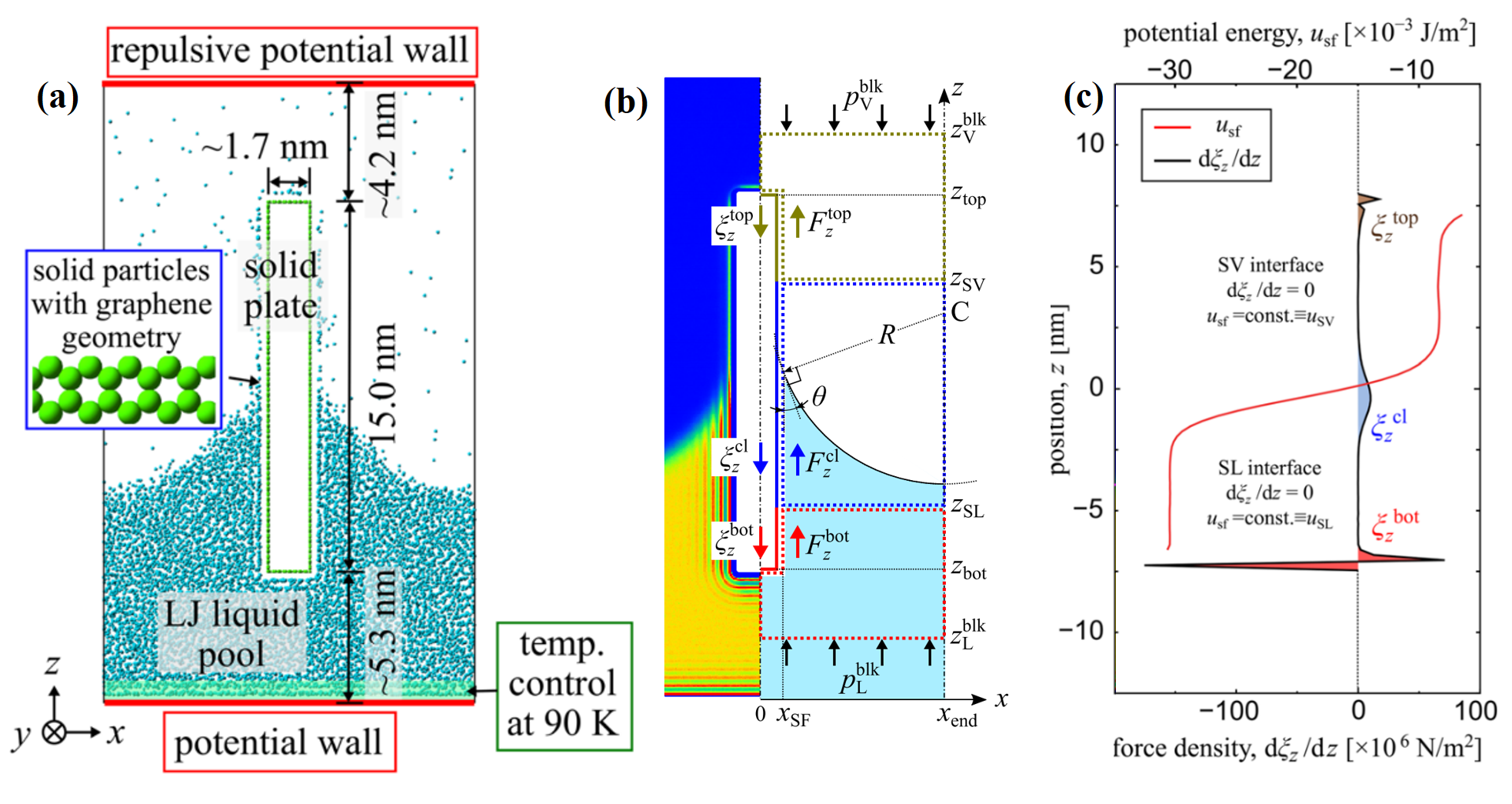}
			\caption{(a) The Wilhelmy simulation system: the molecular system of a quasi-2D meniscus emerges upon a hollow rectangular solid plate, submerged partially in a liquid reservoir.
				(b) The solid substrate has a top, middle (contact line), and bottom parts experiencing downward forces $\xi_z^{\rm top}$, $\xi_z^{\rm cl}$, and $\xi_z^{\rm bot}$ from the fluid, respectively. The control volumes (CVs) of fluids subject to upward forces $F_z^{\rm top}$, $F_z^{\rm cl}$, and $F_z^{\rm bot}$ from the solid.
				(c) Profiles of the force density acting on the solid substrate and solid-fluid (SF) potential energy.
				Figures (a-c) are adapted from Ref.\citenum{imaizumi2020wilhelmy}.}
			\label{fig:z_f6}
		\end{centering}
	\end{figure}

	As shown in Fig. \ref{fig:z_f6}a, the molecular arrangement simulates the experimental configuration, 
	wherein a quasi-2D meniscus emerges upon a hollow rectangular solid plate, submerged partially in a liquid reservoir. 
	The simulation system applies periodic boundary conditions in x- and y-directions. 
	In equilibrium, the system can be approximated as homogeneous in the y-direction (not shown in Fig. \ref{fig:z_f6}a). Two repulsive potential walls were fixed at the top and bottom of the simulation box making the boundary non-periodic in the z-direction. A graphene sheet was bent into a rectangular shape to form the solid plate, which was then positioned in contact with the fluid, and the solid particles were anchored at specific coordinates on this structure. The right and left surfaces of the solid plate were aligned parallel to the yz-plane, while the upper and lower surfaces were oriented parallel to the xy-plane.
	
	Several properties are measured in such molecular systems at an equilibrium state. 
	Most of them are marked in Fig. \ref{fig:z_f6}b. Those properties include: Bulk pressures in the liquid and vapor phases ($P_{\rm V}^{\rm blk}$ and $P_{\rm L}^{\rm blk}$); Contact angle of the meniscus $\theta$; Downward forces $\xi_z^{\rm top}$, $\xi_z^{\rm cl}$, and $\xi_z^{\rm bot}$ experienced by the top, middle (contact line), and bottom parts of the solid plate from the fluid, respectively; Solid-fluid potential energy densities with unit of energy per area $u_{\rm SV}$ and $u_{\rm SL}$ at SV and SL interfacial regions, separately.
	$P_{\rm V}^{\rm blk}$ and $P_{\rm L}^{\rm blk}$ are derived by dividing averaged forces on the top/bottom potential walls by surface areas.
	$\theta$ can be calculated based on equilibrium density distributions (see section 2.1).
	$\xi_z$ ($u_{\rm sf}$) can be computed by summarizing the pair-wise force in z-direction (potential energy) between solid particles per unit area and fluid particles within the cutoff distance.
	Fig. \ref{fig:z_f6}c show typical distributions of $d\xi_z/dz$ and $u_{\rm sf}$, which can be used to extract $\xi_z^{\rm top}$, $\xi_z^{\rm cl}$, $\xi_z^{\rm bot}$, $u_{\rm SL}$ and $u_{\rm SV}$ as noted in the figure. 
	
	The interfacial properties of interest (including $\gamma_{\rm SL}^*, \gamma_{\rm SV}^*, \gamma_{\rm LV}$, and the pinning force at three-phase contact line $F_z^{\rm cl}$ ) can be calculated based on the measured properties using the following relations:\cite{imaizumi2020wilhelmy}
	\begin{equation}
		\label{eq:r1}
		P_{\rm V}^{\rm blk} - P_{\rm L}^{\rm blk} = \cfrac{\gamma_{\rm LV} \cdot {\rm cos} \theta}{x_{\rm end} - x_{\rm SF}},
	\end{equation}
	\begin{equation}
		\label{eq:r2}
		\xi_z^{\rm cl} = F_z^{\rm cl} -  u_{\rm SL} + u_{\rm SV},
	\end{equation}
	\begin{equation}
		\label{eq:r3}
		\xi_z^{\rm bot} = -x_{\rm SF}P_{\rm L}^{\rm blk} - \gamma_{\rm SL}^* + u_{\rm SL},
	\end{equation}
	\begin{equation}
		\label{eq:r4}
		\xi_z^{\rm top} = -x_{\rm SF}P_{\rm V}^{\rm blk} - \gamma_{\rm SV}^* - u_{\rm SV},
	\end{equation}
	where $x_{\rm SF}$ is the position of the solid-fluid interface ($i.e.,$ the nearest location where the fluid can reach near the solid), which can be determined from the density profile. $x_{\rm end}$ is half the size of the simulation box in the x-direction (see Fig. \ref{fig:z_f6}b).
	
	Here, we briefly introduce the relations (Eq. \ref{eq:r1}-\ref{eq:r4}) between interested interfacial properties and the measured properties (full derivations are available in Ref.\citenum{imaizumi2020wilhelmy}). The Eq. \ref{eq:r1} is the Young-Laplace equation.\cite{firoozabadi2016thermodynamics}
	As given in APPENDIX A of the Ref.\citenum{imaizumi2020wilhelmy}, the Eq. \ref{eq:r2} is derived by a mean-field approach through the analysis of forces around the contact line ($i.e.,$ the blue-dotted control volume (CV) shown in Fig. \ref{fig:z_f6}b). In a similar method, the following relation can be derived for the red-dotted CV:\cite{imaizumi2020wilhelmy}
	\begin{equation}
		\label{eq:r5}
		\xi_z^{\rm bot} = F_z^{\rm bot} +  u_{\rm SL},
	\end{equation}
	where $F_z^{\rm bot}$ is the upward forces experienced by the fluid in the red-dotted CV in Fig. \ref{fig:z_f6}b from the solid. In equilibrium, the force balance on the red-dotted CV can be described by:\cite{imaizumi2020wilhelmy}
	\begin{equation}
		\label{eq:r6}
		-\int_{0}^{x_{\rm end}} \tau_{zz}(x,z_{\rm L}^{\rm blk}) dx + \int_{x_{\rm SF}}^{x_{\rm end}} \tau_{zz}(x,z_{\rm SL})dx + F_z^{\rm bot} = 0,
	\end{equation}
	where $\tau_{zz}$ is the principle component of 2D fluid stress tensor (averaged in the y-direction) in the z-direction, and the locations of $z_{\rm L}^{\rm blk}$ and $z_{\rm SL}$ are marked in Fig. \ref{fig:z_f6}b.
	The fluid bulk pressure is related to the fluid stress tensor and the first term in the left-hand side of Eq. \ref{eq:r6} can be expressed as:\cite{imaizumi2020wilhelmy}
	\begin{equation}
		\label{eq:r7}
		-\int_{0}^{x_{\rm end}} \tau_{zz}(x,z_{\rm L}^{\rm blk}) dx = P_{\rm L}^{\rm blk}x_{\rm end}.
	\end{equation}
	Applying Bakker's equation for the SL relative interfacial tension ($i.e.,$ Eq. \ref{eq:bakker2} with shifted origin), we can express the second term in the left-hand side of Eq. \ref{eq:r6} as:\cite{imaizumi2020wilhelmy}
	\begin{equation}
		\label{eq:r8}
		\int_{x_{\rm SF}}^{x_{\rm end}} \tau_{zz}(x,z_{\rm SL})dx = \gamma_{\rm SL}^* - (x_{\rm end} - x_{\rm SF})P_{\rm L}^{\rm blk}.
	\end{equation}
	Substituting Eq. \ref{eq:r5}, \ref{eq:r7} and \ref{eq:r8} into Eq. \ref{eq:r6}, Eq. \ref{eq:r3} is  recovered. By doing the same procedures for the brown-dotted CV, Eq. \ref{eq:r4} can be derived.
	
	To optimize simulation accuracy with the Wilhelmy simulation method, the dimensions of the simulation box and the solid plate need to be carefully calibrated. 
	Firstly, the contact angle $\theta$ was calculated using the same one in the Contact Angle Approach in section 2.1. 
	Therefore, a large size of box length in the x-direction should be used so that the radius of the meniscus is large enough to mitigate the system size effects.\cite{jiang2019recent} 
	Note that the method may also suffer from the hysteresis issue discussed above.\cite{godawat2008structure,macdowell2006nucleation}
	Meanwhile, it is important to ensure that the size of the simulation box in the z-direction is sufficiently large. This allows for the existence of a bulk liquid phase between the solid plate and the potential wall at the bottom. Similarly, it also accommodates the presence of a bulk vapor phase between the solid plate and the potential wall at the top.
	Additionally, the width of the solid plate should be large enough ($i.e.,$ larger than the cutoff distance) to prevent interactions between fluids adjacent to the edges of the plate. Similarly, the length of the plate should be large enough to ensure that the bulk regions of both the liquid and vapor phases are present at the simulation box in the x-directional boundary. This is related to the position of the three-phase contact line which is better positioned near the center of the plate in equilibrium.
	It is also important to note that the derivation assumes that particle interaction potentials are truncated using a cutoff distance. 
	However, studies show that long-range pairwise interactions significantly impact fluid-fluid IFEs.\cite{stephan2020influence}  
	Therefore, a large cutoff distance is recommended, necessitating a solid plate with a large width.
	Consequently, these constraints mandate a large simulation system, entailing considerable computational resources.

	Using the Wilhelmy simulation method, one can obtain the fluid-solid relative IFEs from just one equilibrium molecular simulation, thereby avoiding the complex and often disputed process of calculating local pressure profiles.\cite{shi2023perspective} 
	An additional important advantage of this method is the ability to compute the pinning force. This is particularly useful for contact lines that experience pinning as a result of irregularities, such as surface roughness, impurities, or deformations of the surface.\cite{yamaguchi2019interpretation}
	In the example illustrated above, the solid plate was made using graphene, and
	the derivations employ a mean-field approach, assuming zero-thickness for the solid plate.
	However, the approach can be extended to other solid materials with non-zero thickness. 
	Note that the material for the top and bottom surfaces of the solid plate can be different from that for the left and right surfaces. For example, potential walls could be used for the top/bottom surfaces to reduce the computation cost. 

\begin{figure}[tb]
		\begin{centering}
			\includegraphics[width=0.9\textwidth]{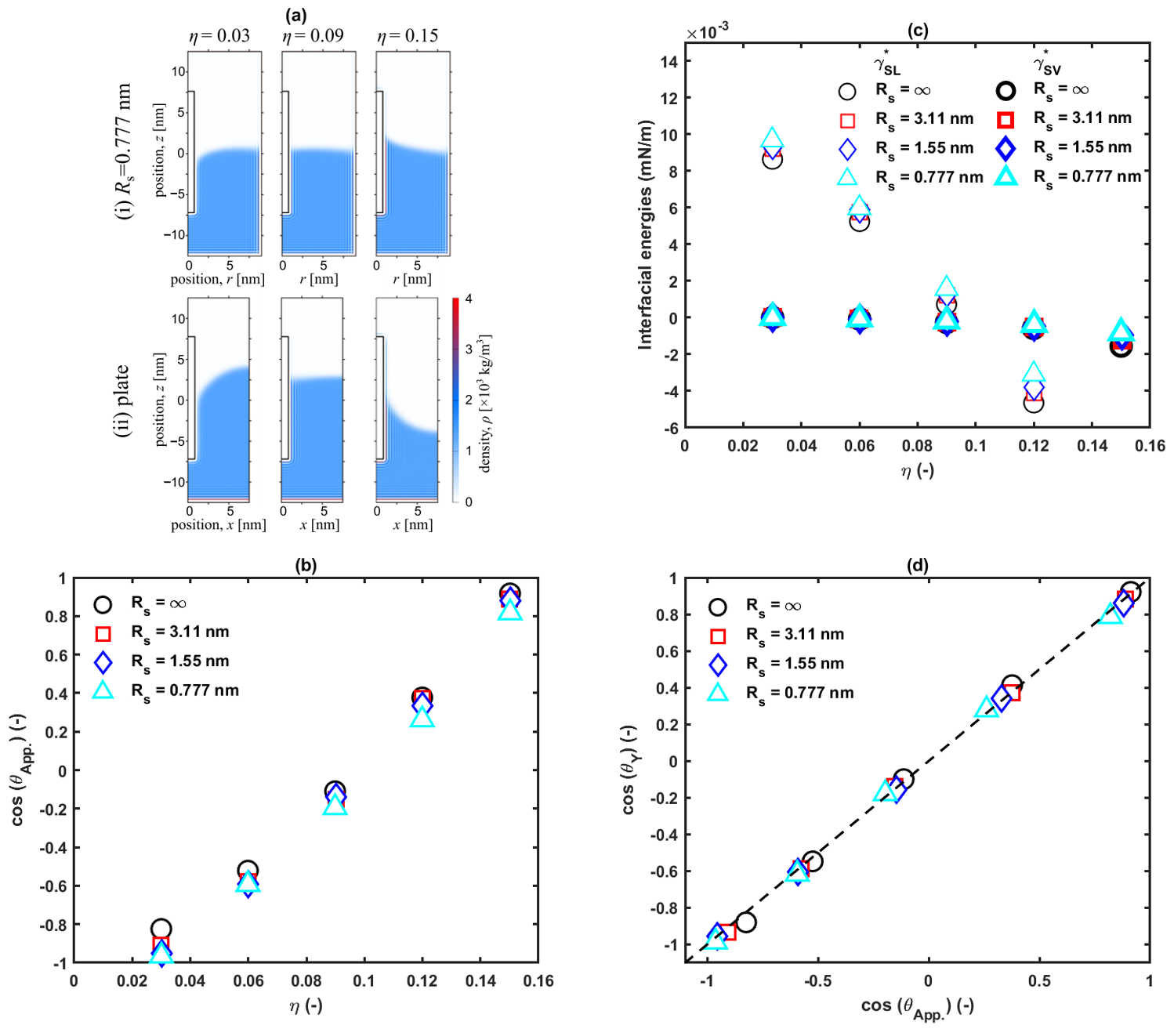}
			\caption{(a) Density profiles around (i) solid cylinder of radius R$_s$ = 0.777 nm and (ii) plate with various fluid-solid interaction parameter $\eta$. 
				(b) Dependence of the wettabilities on $\eta$ for various R$_s$.
				(c) Dependence of the relative IFE of the solid-liquid and solid-vapor interfaces on  $\eta$.
				(d) Comparison between the apparent wettabilities $\theta_{\rm App.}$ and wettabilities from Young's equation $\theta_{\rm Y}$ using the relative IFEs of the fluid-solid interfaces and fluid-fluid IFE with varying $\eta$ and R$_s$.
				Figure (a) is adapted from Ref.\citenum{watanabe2022quantifying}. The data for figures (b-d) are taken from Ref.\citenum{watanabe2022quantifying}.}
			\label{fig:z_f7}
		\end{centering}
	\end{figure}

	Remarkably, the method has been extended to deal with cylindrical-shaped solid plates to understand the effects of curvature on fluid-solid interfaces.\cite{watanabe2022quantifying}
	The interfacial characteristics of a Lennard-Jones fluid around solid cylinders with different radii are displayed in Fig. \ref{fig:z_f7}.
	Fig. \ref{fig:z_f7}a displays the density distributions surrounding the plate and the solid cylinder with the smallest radius, $R_s$ = 0.777 nm.
	It is intriguing to note that the cylinder and plate appear to have substantially distinct meniscus forms. This implies that in order to accurately estimate the wettabilities from the meniscus shape, different force balances should be used.\cite{watanabe2022quantifying}
	The wettability was determined by the fluid-solid interaction parameter $\eta$. $\eta$ is implemented in the pair-wise potential energy between the fluid particle and the solid particle: $\varepsilon_{\rm ij} = \eta \cdot \varepsilon_{\rm ij}^{\rm Mix}$, where $\varepsilon_{\rm ij}^{\rm Mix}$ is the potential well depth obtained from the mixing rule.\cite{watanabe2022quantifying}
	The contact angle $\theta_{\rm App.}$ derived from the meniscus forms decreases as $\eta$ increases (see Figs. \ref{fig:z_f7}a and b). Furthermore, a decreased radius is observed to result in a slight rise in the contact angle.
	The relative IFEs of the liquid-solid and vapor-solid interfaces are displayed in Fig. \ref{fig:z_f7}c. The effects of fluid-solid interaction on relative IFE are similar to those in water/solid system\cite{yang2023molecular} discussed above (c.f. Figs. \ref{fig:z_f5}a,b and Fig. \ref{fig:z_f7}c). It is important to note that reducing the radius leads to an increase in the relative IFE values.
	In order to validate the calculated relative IFEs, those data were further substituted into Young's equation to estimate the contact angle $\theta_{\rm Y}$. As seen in Fig. \ref{fig:z_f7}d, this contact angle matches perfectly with $\theta_{\rm App.}$ obtained from fitting the meniscus shape.\cite{watanabe2022quantifying}
	
	A potential extension of the Wilhelmy method includes handling the solid in a spherical shape. 
	On the other hand, advancing this method to incorporate more complex systems represents a pivotal aim for forthcoming research. For example, the method could be applied to the system in the presence of electrostatic interactions within solid and fluid molecules.
	%
	
	\section{Thermodynamic Routes}
\begin{figure}[tb]
		\begin{centering}
			\includegraphics[width=0.9\textwidth]{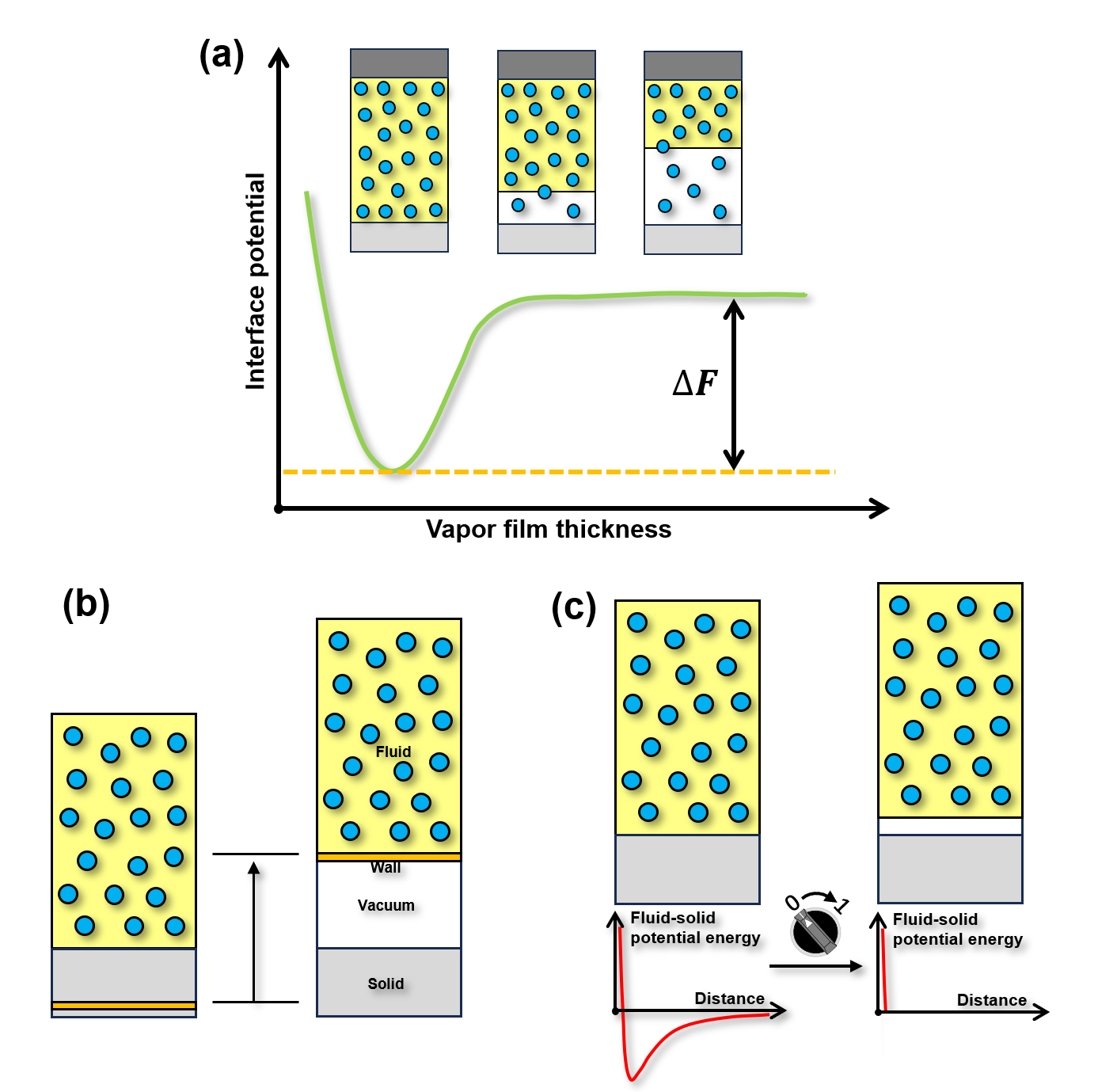}
			\caption{Sketches of TI paths: (a) the interface potential method developed by Errington and co-workers\cite{rane2011monte,kumar2014use}, (b) the phantom-wall method developed by Leroy et al.\cite{leroy2009interfacial}, and (c) the dry-surface method developed by Leroy and M\"{u}ller-Plathe.\cite{leroy2015dry}
				The blue particles in the yellow regions denote the fluid. The white, grey, dark grey, and orange regions denote the vacuum, solid, sticky solid, and phantom-wall, separately. 
				Figures (a), (b), and (c) are adapted from Refs.\citenum{rane2011monte}, \citenum{leroy2009interfacial}, and \citenum{leroy2015dry}, respectively.}
			\label{fig:z_f8}
		\end{centering}
	\end{figure}

	The thermodynamic route refers to a series of methods based on thermodynamic principles to calculate IFE in fluid-solid systems. These methods involve monitoring changes in free energy with respect to interface area in molecular simulations.
	In the review paper of Jiang and Patel\cite{jiang2019recent}, several methods of the thermodynamic route that can be used to calculate the superficial tension and the fluid-solid relative IFE are discussed in detail. Here, we briefly discuss those methods with recent advances for completeness. 
	Errington and co-workers\cite{rane2011monte,kumar2014use} developed the interface potential methods. 
	By systematically changing the chemical potential of the fluid in a grand canonical Monte Carlo simulation, a fluid-solid interface is gradually changed from a liquid-solid to a vapor-solid interface during a drying process or in a reverse direction during a wetting process. A sketch of the method is given in Fig. \ref{fig:z_f8}a. 
	This method calculates superficial tension based on the thickness of the vapor/liquid film, using the grand canonical transition matrix Monte Carlo algorithms.\cite{errington2003direct,grzelak2008computation,grzelak2010calculation} 
	From this, the fluid-solid relative IFE can also be derived.
	Recent developments of this method involve the implementations within an isothermal-isobaric ensemble using Monte Carlo simulation methods\cite{jain2019using,jain2019application} and canonical molecular dynamics simulations.\cite{jain2019construction} 
	
	The relative IFEs of the fluid-solid interface could also be estimated through a thermodynamic integration (TI) scheme using molecular dynamics simulations.
	Leroy et al.\cite{leroy2009interfacial} developed the phantom-wall method. In this method, a repulsive wall with interactions only with the fluid (by turning off the wall-solid interactions) is reversibly shifted away from the surface. This converts the fluid-solid interface into a vacuum-solid interface and a repulsive wall-fluid interface.
	The sketch of such a process has been presented in Fig. \ref{fig:z_f8}b.  
	The relative IFE could be calculated by subtracting the relative IFE of the repulsive wall-fluid interface from the free energy change of the separation process. It is important to note that the relative IFE of the repulsive wall-fluid interface is typically approximated by the fluid interfacial tension. This approximation has been demonstrated to be sufficiently accurate in scenarios involving water near purely repulsive walls at room temperature.\cite{leroy2011rationalization,huang2001scaling}  However, the error of this approximation could be significant if the effects of the vapor phase on the interfacial tension cannot be disregarded, such as in the case of water under high temperatures.\cite{leroy2015dry} In such instances, the technique utilizing Bakker's equation may be a more suitable choice for estimating the relative IFE.
	%
	Recently, Uranagase et al.\cite{uranagase2018efficient} developed a scheme for calculating the work of adhesion between a liquid and complex surface, and the scheme could also be used to estimate the fluid-solid  relative IFE. Such a scheme has been implemented in a freely available code named ``FE-CLIP" by Uranagase and Ogata.\cite{uranagase2020fe}

	The dry-surface method, developed by Leroy and M\"{u}ller-Plathe, utilizes a procedure in which the attractions between the surface and the fluid are progressively turned off (see sketch shown in Fig. \ref{fig:z_f8}c).\cite{leroy2015dry} 
	Recently, Surblys et al.\cite{surblys2022computing} addressed the challenge of long-range Coulombic interactions at interfaces in the dry-surface method by substituting them with damped Coulomb interactions and investigating various thermal integration paths.
	It is worth mentioning that the dry-surface method has been used as the reference method for validating the Wilhelmy simulation method for surfaces with various wettabilities.\cite{imaizumi2020wilhelmy} 
	However, similar to the potential issue faced in the phantom-wall method, estimating the relative IFE within the dry-surface method also operates under the assumption that the relative IFE of the fluid-repulsive wall interface can be approximated as the fluid interfacial tension.
	Additionally, Kandu{\v{c}} and Netz\cite{kanduvc2015hydration,kanduvc2017atomistic} have introduced a TI scheme that is anticipated to be more effective in analyzing the wetting properties of highly hydrophilic surfaces.
	
	In the next sections, we discuss other important methods of the thermodynamic route for estimating fluid-solid IFE, which are classified into two groups, namely, TI methods and free energy perturbation (FEP) methods. 
	
	\subsection{Thermodynamic Integration Methods}
	
	As noted above, TI methods in molecular simulations are used to calculate free energy differences between two states by systematically integrating energy changes during state transitions.\cite{frenkel2023understanding} 
	The literature on TI methods developed for computing the fluid-solid IFE is numerous. 
	In addition to the methods mentioned above, here, we list several other methods including the Gibbs-Cahn TI techniques,\cite{frolov2009temperature,laird2009determination,frolov2009solid} the ensemble mixing/switch methods,\cite{heni1999interfacial,deb2011hard,deb2012methods,schmitz2015ensemble,virnau2016ensemble} 
	the Schilling-Schmid method,\cite{schilling2009computing,schmid2010method}
	the method using the Gibbs adsorption equation,\cite{mao1997density} the method based on the excess grand potential of confined fluids\cite{hamada2007phase}, and the method of Das and Binder for superficial tensions.\cite{das2010does,das2011simulation}
	In the following parts, we focus on two typical TI methods for computing fluid-solid IFEs, namely, the cleaving wall method and the Frenkel-Ladd technique.
	
	\subsubsection{Cleaving Wall Method}

	%
	In this section, we begin with a concise overview of developments in cleaving wall methods. We provide a detailed discussion of two distinct versions of these methods: one for calculating the fluid-solid IFE, $\gamma_{\rm SF}$, and the other for the relative IFE, $\gamma_{\rm SF}^*$. Finally, a practical application example will be presented.
	Note that the cleaving wall method is also capable of calculating the vacuum-solid IFE $\gamma_{\rm S}$ directly using a similar procedure.\cite{broughton1986molecular,tipeev2021direct}
	
	The cleaving wall method employs a reversible process to separate a fluid-solid system into distinct bulk fluid and solid phases using external potential walls. 
	The methods share common features with the phantom-wall method mentioned above, and the latter can be considered as a special case of the former.
	Eriksson\cite{eriksson1969thermodynamics}  introduced the concept of ``cleaving", which has since evolved,\cite{cammarata1994surface,haiss2001surface} despite early debates about the reversibility of the process.\cite{marichev2009current,marichev2011shuttleworth}
	The rigorous definition of the concept has been built based on statistical mechanics.\cite{chipot2007free}
	The use of reversible cleaving in molecular simulations began with Miyazaki et al.\cite{miyazaki1976new} 
	Then, Broughton and Gilmer developed the cleaving wall method in molecular dynamics for creating an interface in a liquid-solid system.\cite{broughton1986molecular}
	Their approach employed uniquely designed ``cleaving potentials" to divide bulk liquid and solid phases into two segments separated by a cleaving plane, followed by merging these segments and ultimately eliminating the ``cleaving potential". 
	However, the method, which involves creating specific cleaving-potential functions for each crystal facet, faces challenges due to the high uncertainty from hysteresis and the complexity of designing these potentials for various orientations and conditions.
	The approach was further developed by Davidchack and Laird.\cite{davidchack2000direct,davidchack2003direct,davidchack2005crystal,laird2005direct,davidchack2010hard}
	To resolve the anisotropy in IFE, they adopted flat cleaving potentials constructed from particles similar to those in the system, comprising several crystalline layers that mirror the structure of the actual crystal phase.
	However, eliminating the hysteresis caused by the movement of the crystal-liquid interface appears to be a challenging issue to fully resolve.
	Nevertheless, attempts have been made to address this problem by carrying out several independent TI runs and choosing the one that demonstrated the least amount of hysteresis.\cite{davidchack2003direct,davidchack2010hard}
	
	Benjamin and Horbach\cite{benjamin2014crystal} proposed a version of the cleaving wall method to address the hysteresis issue by utilizing a planar wall with an extremely short interaction range (modeled using a Gaussian potential) to distinctly separate the liquid and solid phases.
	Here we briefly illustrate the TI path for the calculation of the fluid-solid IFE within this method. 
	The TI scheme starts with two separate simulations: one of a bulk crystal and the other of a bulk liquid, both conducted under coexistence conditions and ends with a combined crystal-liquid two-phase system containing two fluid-solid interfaces. 
	
\begin{figure}[tb]
		\begin{centering}
			\includegraphics[width=0.9\textwidth]{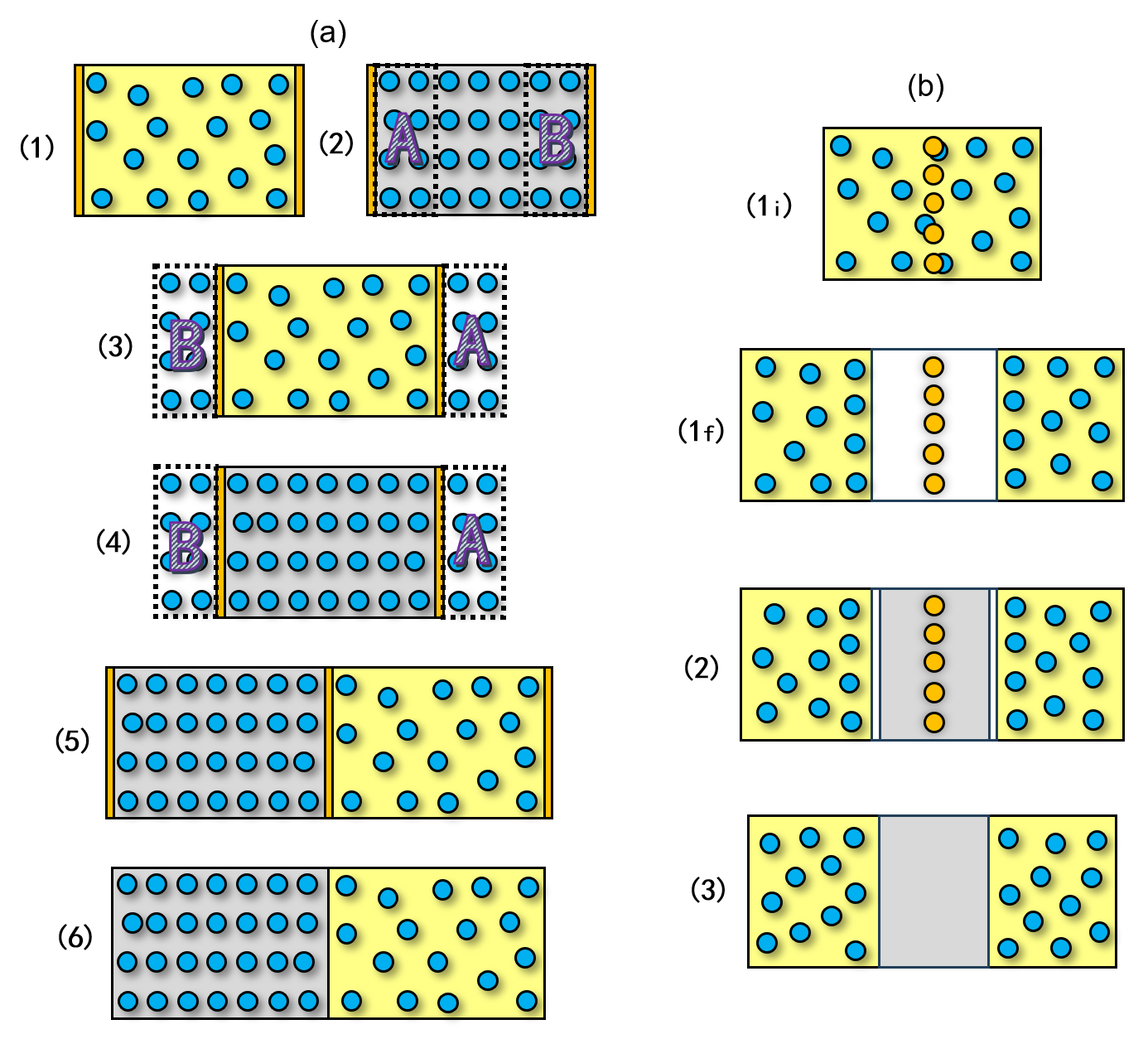}
			\caption{Sketches of TI paths: (a) The Benjamin and Horbach version of the cleaving wall method for IFE,\cite{benjamin2014crystal} and 
				(b) the Addula and Punnathanam version of the cleaving wall method for relative IFE.\cite{benjamin2014crystal} Subscripts i and f stand for initial and final state, respectively.
				The blue particles in the grey and yellow regions denote the solid and the liquid, separately. The orange regions in (a) denote very short-ranged Gaussian walls. The orange particles in (b) are the wall atoms. 
				Figures (a) and (b) are adapted from Refs.\citenum{benjamin2014crystal} and \citenum{addula2020computation}, respectively.}
			\label{fig:z_f9}
		\end{centering}
	\end{figure}

	The TI scheme, which includes 6 steps as depicted in Fig. \ref{fig:z_f9}a, should be carried out progressively through each step.
	Step 1: A flat potential wall, with a short interaction range modeled by a Gaussian function, is placed at the end of the simulation box along the z-axis.
	Step 2: Repeat step 1 for the simulation box of the bulk crystal.
	Step 3: Create two solid walls (shown as A and B inside grey dotted boxes in sketch (2)).  
	The solid walls consist of 2 to 3 layers of crystalline particles adjacent to the Gaussian wall that has been inserted in step 2.
	Subsequently, these walls are connected to the correct ends of the liquid simulation box from step 1. Meanwhile, the periodic boundary conditions in the z-direction of the liquid simulation cell are deactivated.
	Step 4: Repeat step 3 for the crystal simulation box from step 2.
	Step 5: Merge the liquid and solid systems obtained from Steps 3 and 4. 
	Activate the interactions between two phases at the contact, and also across the box boundary by enabling the periodic boundary condition in the z-direction.
	Step 6: Remove the very short-ranged Gaussian walls.
	
	It is crucial to mark that the Gaussian wall serves solely to stop the liquid and crystalline particles from passing the border without altering the bulk properties. 
	The usage of the very short-range Gaussian wall necessitates a very small timestep in molecular dynamics simulations. 
	However, this difficulty can be effectively addressed by employing a multiple time step algorithm.\cite{frenkel2023understanding}
	Moreover, the incorporation of structured solid walls in steps 3 and 4 aims to minimize the disruption of the crystal during the merge of the two phases. 
	This is vital to ensure that the liquid organizes into layers in the interfacial region, aligning with the interactions from the crystal.
	Remarkably, as the barrier of the Gaussian wall weakens in step 6, particles gain the ability to traverse the boundaries, potentially causing the interfaces to shift and resulting in hysteresis along the TI path. Nevertheless, the approach effectively addresses this hysteresis due to the minimal impact of the final step on fluid-solid IFE as a result of the extremely short-ranged flat walls.
	
	The above-mentioned cleaving wall methods involve manipulating the periodic boundary conditions ($e.g.,$ steps 3-5 in Fig. \ref{fig:z_f9}a). Such a function requires specialized implementation and is not available in most of the open-source codes. Moreover, analytical differentiation of the potential energy can be challenging to calculate. To generalize the application of the Benjamin and Horbach approach, Qi et al.\cite{qi2016obtaining} proposed a multi-scheme TI method combined with numerical approximation of thermodynamic integrands to circumvent these problems. 
	The scheme has been utilized for determining the IFEs of the Ag-ethylene glycol interface and holds potential for application across a wide variety of systems of interest.
	It is noteworthy that a LAMMPS package ``CLEAVING" has been recently released for the calculation of IFE of solid-fluid and solid-solid interfaces using the cleaving wall method.\cite{di2024cleaving}  This package enables the reproduction of results from Refs. \citenum{di2022cleaving},\citenum{di2020shuttleworth}, and \citenum{davidchack2003direct}.
	%
	
	%
	Building on prior methods, Addula and Punnathanam\cite{addula2020computation}  developed a cleaving wall technique that exclusively cleaves the fluid phase.
	This method integrates the Frenkel-Ladd approach (outlined in section 3.1.2) to calculate the vacuum-solid IFE. The overall IFE can be estimated by combining these two components.
	As shown in Fig. \ref{fig:z_f9}b, this version of the cleaving wall method includes 3 steps.
	In the first step, the bulk liquid is split by applying a cleaving potential to create a vacuum region. The initial and final snapshots of the first step are shown in Fig. \ref{fig:z_f9}b(1i) and (1f), respectively. 
	Note that a cleaving potential is composed of atoms that exclusively interact with the fluid, incorporating both repulsive and attractive terms.
	In the second step, the solid slab is placed into the created region (see Fig. \ref{fig:z_f9}b(2)). In the final step, the cleaving potential is withdrawn, leading to the development of two  solid-liquid interfaces  (see Fig. \ref{fig:z_f9}b(3)).
	All procedures are performed within the NP$_z$T ensemble.\cite{pedersen2013computing} The free energy changes in step 1 ($\Delta \rm G_1$) and step 3 ($\Delta \rm G_3$) are determined through thermodynamic integration. For step 2, the free energy change ($\Delta \rm G_2$) is estimated using the thermodynamic perturbation method.
	
\begin{figure}[tb]
		\begin{centering}
			\includegraphics[width=0.8\textwidth]{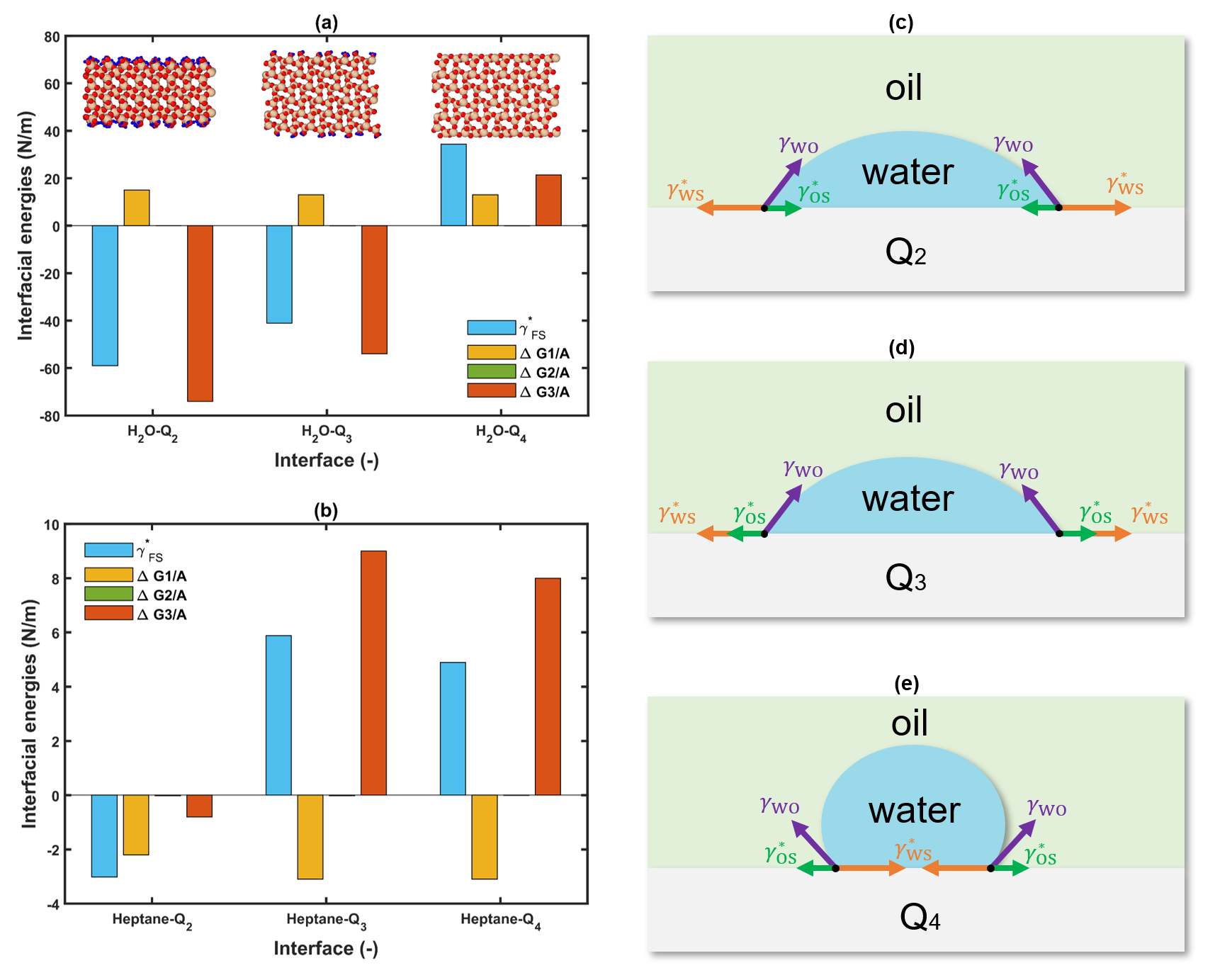}
			\caption{The relative IFEs and contributions from each TI step for the (a) water/silica and (b) the oil (heptane)/silica interfaces with various silanol group densities. The insets of figure (a) display the Q$_2$, Q$_3$, and Q$_4$ silica surface models, arranged from left to right. The color code is as follows: Si (brown), O (red), and H (blue).
				The interfacial energy data are taken from Ref.\citenum{patel2022computing} and silica models are visualized with atom coordinates from Ref.\citenum{emami2014force}.
				(d), (e), and (f) show directions of interfacial energies on Q$_2$, Q$_3$, and Q$_4$ surfaces during the wetting/dewetting processes before the equilibrium, respectively.		
				$\gamma_{\rm WO}$, $\gamma_{\rm WS}^*$, and $\gamma_{\rm OS}^*$ denote water-oil IFE, water-solid relative IFE, and oil-solid relative IFE, respectively.}
			\label{fig:z_f10}
		\end{centering}
	\end{figure}

	The method has an advantage over the phantom wall method in that it eliminates the need to estimate the fluid-wall IFE. Nevertheless, it is more complex, as it requires three thermodynamic integration steps, whereas the phantom wall method necessitates only one. 
	Additionally, it is important to note that a strong cleaving potential can induce phase transitions, leading to an overly organized fluid structure near the wall.\cite{leroy2010solid,addula2020computation}
	Therefore, calibrating the interaction strength of the cleaving potential is crucial to maintaining a reversible path and preventing such phase transitions that violate the reversibility condition.
	The cleaving wall method of Addula and Punnathanam\cite{addula2020computation} has been applied to investigate the wettabilities of water/oil/rock systems.\cite{patel2022computing} 
	Understanding the wettability of water/oil/rock systems is crucial for optimizing enhanced oil recovery processes.\cite{morrow1990wettability,ding2021pore}
	Fig. \ref{fig:z_f10}a and b present relative IFEs and their contributions from each TI step of the water/solid and water/oil interfaces, respectively.
	Three silica surfaces with different hydrophilic silanol densities are studied, namely, Q$_2$ with Si-OH density of 9.4 nm$^2$ ,Q$_3$ with Si-OH density of 4.7 nm$^2$, and Q$_4$ with Si-OH density of 0 nm$^2$ (see snapshots in Fig. \ref{fig:z_f10}a).\cite{emami2014force}
	Notably, the free energy change from the insertion of the solid slab $\Delta \rm G_2$ ($i.e.,$, step 2) is little in contrast to those in other steps. This is attributed to the strong attraction between the cleaving wall atoms and the fluid, which preserves the structure of the fluid phase during insertion.\cite{patel2022computing}
	
	As the silanol density decreases ($i.e.,$, from Q$_2$ to Q$_4$), the relative IFEs change from negative to positive, which is consistent with the results given in Fig. \ref{fig:z_f5} and Fig. \ref{fig:z_f7}. 
	The magnitudes of relative IFEs of water-silica interfaces are much larger than those of heptane-silica interfaces.
	For the water-silica interface, both Q$_2$ and Q$_3$ are water-wet surfaces, while Q$_4$ is hydrophobic in water/vacuum/silica systems by comparing Fig. \ref{fig:z_f10}a and Fig. \ref{fig:z_f3}.
	For the heptane-silica interface, only Q$_2$ surface is oil-wet, while oil contact angles in the water/vacuum/silica (Q$_3$ and Q$_4$) systems are greater than 90$\degree$ by comparing Fig. \ref{fig:z_f10}b and Fig. \ref{fig:z_f3}. 
	Since the miscibility of water and oil is low,\cite{maczynski2004recommended} those relative IFEs can be substituted into Young's equation to estimate the wettability. The water contact angles are 0$\degree$, 0$\degree$, and 128$\degree$ for Q$_2$, Q$_3$, and Q$_4$ surfaces, respectively.\cite{patel2022computing} Those values are in excellent agreement with results obtained from the contact angle approach.\cite{patel2022computing}
	Furthermore, the directions of interfacial energies during the wetting and dewetting processes are illustrated in Figs. \ref{fig:z_f10}c-e. 
	Notably, the relative IFEs for both water-solid and oil-solid interfaces align in the same direction on the Q$_3$ surface, promoting the spreading of the water droplet.

	\subsubsection{Frenkel-Ladd Technique}
	
	The Frenkel-Ladd technique, introduced in 1984, provides a method for calculating the absolute Helmholtz free energy of periodic crystals.\cite{frenkel1984new} 
	The method establishes a reversible thermodynamic pathway connecting the solid with an Einstein crystal, the free energy of which can be analytically determined.
	Note that in an Einstein crystal, each atom is confined within its own harmonic potential well, which isolates them from interacting with one another.\cite{einstein1907plancksche} 
	In other words, the potential energy is solely determined by the positions of the atoms within their respective harmonic potential wells, and the potential energy of the system is independent of the relative positions of the atoms.
	The developments and applications of the Frenkel-Ladd technique for free energies of bulk solids have been summarized in several review papers.\cite{monson2000solid,vega2008determination,reddy2021review}
	
	Pretti and Mittal\cite{pretti2019extension} extended the Frenkel-Ladd technique to non-periodic and semi-periodic systems. This advancement enables the determination of absolute free energies in finite-sized crystals characterized by distinct shapes and surface structures in contact with a vacuum environment ($e.g.,$ $\gamma_{\rm S}$).
	Addula and Punnathanam\cite{addula2020computation} proposed two TI methods for computing the fluid-solid IFE $\gamma_{\rm SF}$.
	The first TI method is based on the cleaving wall method discussed in section 3.1.1 (also see Fig. \ref{fig:z_f9}b). 
	The second approach, known as the ``adsorption method," involves the integration of an adsorption isotherm.
	Both of these methods calculate the fluid-solid relative IFE $\gamma_{\rm SF}^*$.
	And the solid-vacuum IFE ($i.e.,$ $\gamma_{\rm S}$) is handled by a modified Frenkel-Ladd technique.\cite{reddy2018calculation}
	The fluid-solid IFE $\gamma_{\rm SF}$ can then be derived by combining $\gamma_{\rm SF}^*$ and $\gamma_{\rm S}$ (see Eq. \ref{eq:RSF}).
	
\begin{figure}[tb]
		\begin{centering}
			\includegraphics[width=0.9\textwidth]{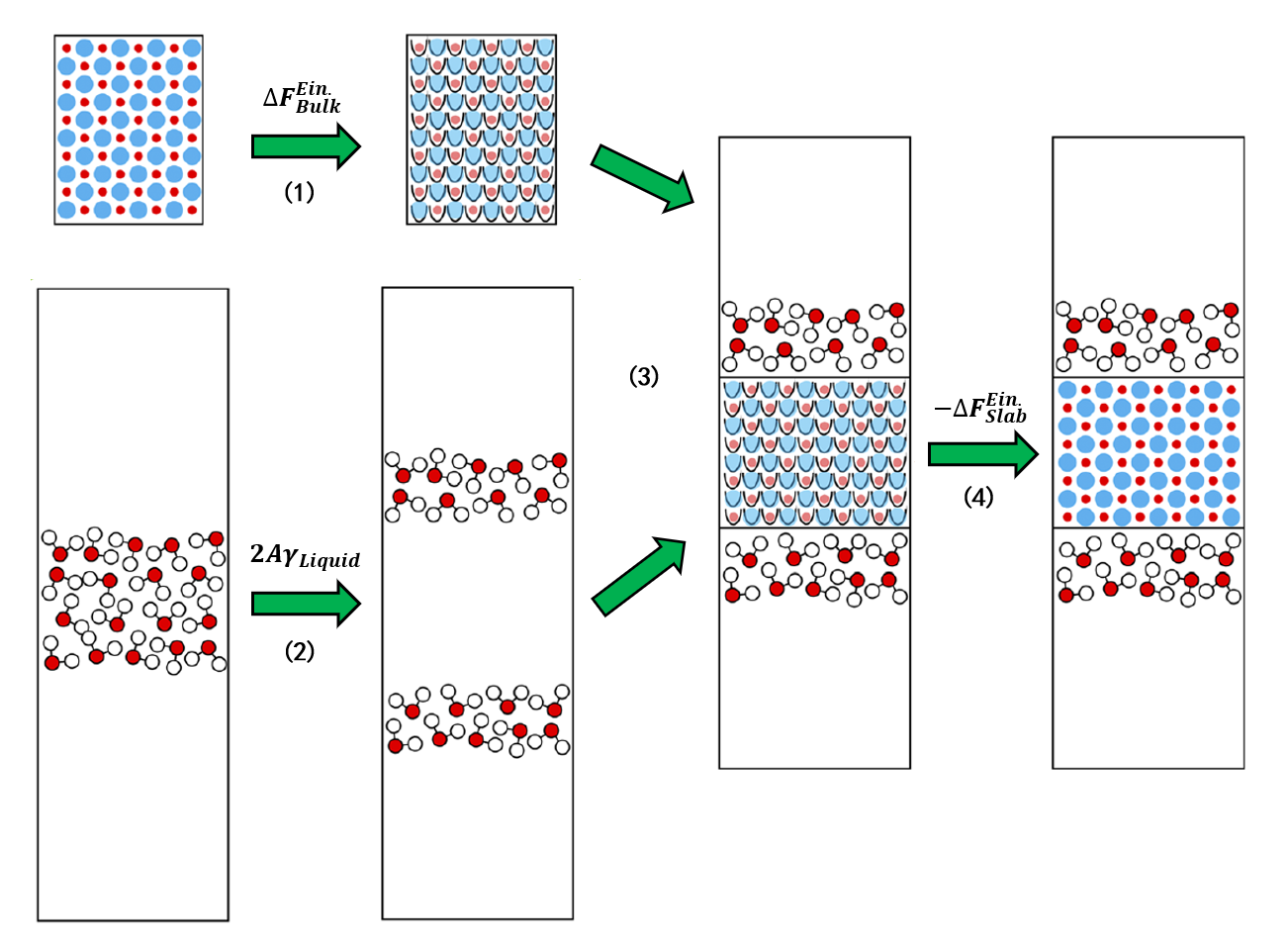}
			\caption{Sketches of TI path for the Frenkel-Ladd technique developed by Yeandel et al.\cite{yeandel2022general} 
				The water molecule is depicted as one red circle (oxygen atom) connected to two white circles (hydrogen atoms).
				The non-connecting blue and red circles denote the solid atoms and the corresponding atoms in the Einstein crystal are marked by the harmonic wells.
				The figure is adapted from Ref.\citenum{yeandel2022general}. }
			\label{fig:z_f11}
		\end{centering}
	\end{figure}

	Recently, Yeandel et al.\cite{yeandel2022general} presented a similar but more general method for computing $\gamma_{\rm SF}$. The method also uses the Einstein crystal as the reference state. 
	The following describes the four-step process in the TI path, illustrated in Fig \ref{fig:z_f11}:
	
	Step (1) transforms the bulk solid into an Einstein crystal through two stages: 
	1. Turn on the harmonic wells for every atom; 2. Turn off all interactions both between and within molecules. The corresponding free energy change is denoted as $\Delta F_{Bulk}^{Ein.}$.
	
	Step (2) creates a vacuum space inside a liquid film. The free energy change of this step is simply $2 A \gamma_{\rm Liquid}$, here $\gamma_{\rm Liquid}$ is the vacuum-liquid IFE. 
	This term can be approximated as the vapor-liquid IFE and calculated using the relatively efficient method based on Bakker's equation (see Section 2.1).
	
	Step (3) inserts the Einstein crystal from step (1) into the middle vacuum space of the split liquid film from step (2). Since there are no interactions between the Einstein crystal and the liquid, the free energy change in this step is zero.
	
	Step (4) turns the liquid-Einstein crystal system from step (3) into the liquid-slab solid system. 
	The corresponding free energy change is denoted as $\Delta F_{Ein.}^{Slab}$. 
	$\Delta F_{Ein.}^{Slab}$ is calculated reversely noting $\Delta F_{Ein.}^{Slab} = -\Delta F_{Slab}^{Ein.}$.
	Two TI stages are used starting from the liquid-slab solid system: 1. Turn on the harmonic wells for every atom in the solid slab; 2. Turn off all the solid-liquid and solid-solid interactions at the same time.
	
	Finally, the fluid-solid IFE can be calculated as follows:\cite{yeandel2022general}
	\begin{equation}
		\label{eq:FLM}
		\gamma_{\rm SF} = \gamma_{Liquid} + \cfrac{\Delta F_{Bulk}^{Ein.}-\Delta F_{Ein.}^{Slab}}{2A}.
	\end{equation}
	
	It is important to note that an assumption of zero vacuum-vapor IFE is made in step (2). This approximation reduces computation cost because the vapor-liquid IFE can be calculated by the efficient method based on Bakker's equation. However, when the vacuum-vapor IFE significantly deviates from zero ($e.g.,$ cases with high vapor density), the cleaving wall method of Addula and Punnathanam\cite{addula2020computation} could be applied for computing $\gamma_{Liquid}$ instead of the method based on Bakker's equation. 
	Moreover, liquid molecules may move through the solid part when transforming an immersed slab into an Einstein crystal in step (4), potentially causing high forces and instabilities that may crash the simulation.
	To avoid this situation, two extra TI stages can be introduced to add and remove potential walls, protecting the slab during its transformation, as implemented by Benjamin and Horbach in their cleaving wall method.\cite{benjamin2014crystal}
	
	One key advantage of this method in contrast to methods of Addula and Punnathanam\cite{addula2020computation} is that it incorporates correction for miscible species, which is particularly useful for some solid substances containing miscible components that are loosely attached to the surface and can move into the liquid layer. The procedures for calculation of this correction are detailed in the original reference\cite{yeandel2022general} and are not included here.
	In addition, the inclusion of vacuum regions on both sides of the liquid-solid-liquid setup (see step (4) of Fig. \ref{fig:z_f11}) enables the method to study surfaces with dipole moments by applying the dipole correction method of Ballenegger et al.\cite{ballenegger2014communication,ballenegger2009simulations}.
	The LAMMPS scripts necessary for implementing this approach are provided in the supplementary material of the Ref. \citenum{yeandel2022general}.
	These features make the method highly promising for applications across a wide range of realistic systems.
	
	%
	The approach has been utilized to examine IFEs of the calcium sulfate hydrate systems to demonstrate the advantages of the method.\cite{yeandel2022general} Calcium sulfate hydrate exists in three distinct hydration phases: anhydrous anhydrite (CaSO$_4\cdot$0H$_2$O), hemihydrate bassanite (CaSO$_4\cdot$0.5H$_2$O), and dihydrate gypsum (CaSO$_4$$\cdot$2H$_2$O). Due to the presence of strongly binding Ca$^{2+}$ ions, significant ordering of the water layer on the interfaces is anticipated. Additionally, the structure contains stoichiometric water, indicating that interfacial water molecules are formally part of the solid, yet exhibit liquid-like behavior. Therefore, when estimating the IFE, it is important to include the correction term for the miscible species.\cite{yeandel2022general}
	
\begin{figure}[tb]
		\begin{centering}
			\includegraphics[width=0.65\textwidth]{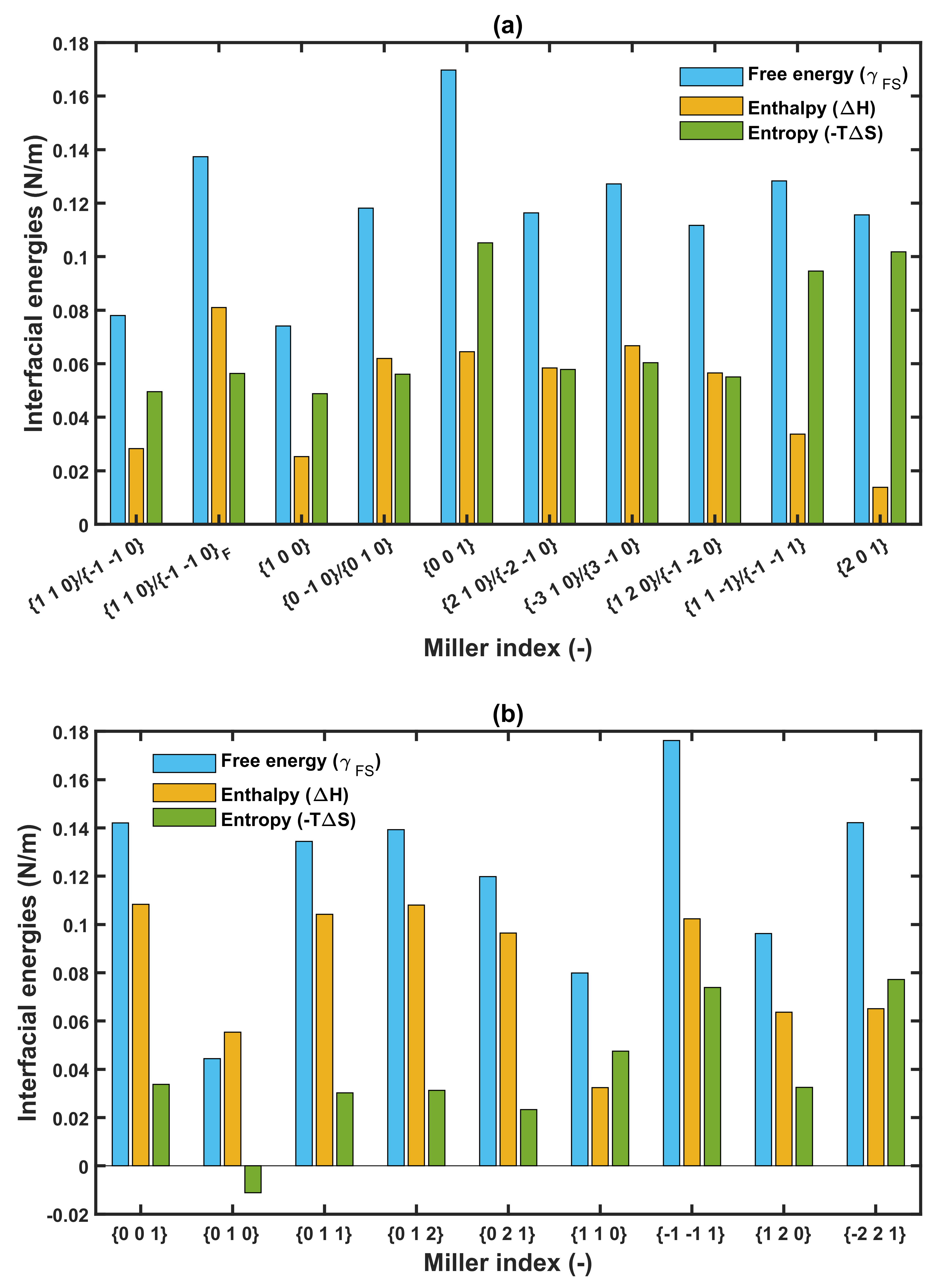}
			\caption{Interfacial energies of (a) bassanite/water and (b) gypsum/water interfaces with different Miller indexes. 
				IFEs $\gamma_{\rm FS}$ are calculated with Eq. \ref{eq:FLM} with additional correction term for miscible species, and enthalpies are calculated by subtracting the potential energy density of each bulk phase and the density of the liquid/vacuum interfacial enthalpy from the potential energy density of the slab system. 
				The entropies are estimated by subtracting the enthalpies from the IFE.
				The data are taken from Ref.\citenum{yeandel2022general}.}
			\label{fig:z_f12}
		\end{centering}
	\end{figure}

	Fig. \ref{fig:z_f12}a illustrates the IFEs of nine bassanite interfaces with varying Miller indices in contact with water. Both the enthalpic and entropic contributions were found to be positive. The enthalpic contribution is typically minimal due to the strong interaction between water and Ca$^{2+}$ ions, which lowers the energy expanse for creating the interface.\cite{yeandel2022general} Conversely, the strong water binding causes significant ordering of interfacial water, leading to an entropy loss compared to the bulk water ($i.e.,$, negative $\Delta$S) and a substantial entropic destabilization.\cite{yeandel2022general} 
	Entropic contributions to the IFE range from around 40\% to 90\%.
	Additionally, two variants of the bassanite $\{$1 1 0$\}$ interface were analyzed. The first variant is highly rough, featuring interface crenellations of CaSO$_4$ chains. The second variant, indicated by ``F" in the subscript, has these crenellations removed. It was found that the crenelated interface is significantly more stable, mainly due to smaller enthalpic contributions. 
	This increased stability is due to the enhanced surface area provided by CaSO$_4$ crenellations, which strengthens water binding.\cite{yeandel2022general} 
	Notably, the entropy contributions to the free energies are nearly identical for both interfaces, implying similar water ordering on each. This similarity in water ordering was explained by considering the arrangement of water molecules beyond the initial adsorbed layer.\cite{yeandel2022general}
	
	The IFEs and IFE contributions of the gypsum interfaces are shown in Fig. \ref{fig:z_f12}b. Generally, the IFEs of gypsum interfaces are quite similar to those of bassanite. 
	But the $\{$0 1 0$\}$ interface exhibits a negative entropy contribution to the IFE ($i.e.,$, positive $\Delta$S). This was explained by the highly structured water layers within the gypsum structure, which become exposed when the [0 1 0] plane is cleaved.\cite{yeandel2022general} 
	As these water molecules gain disorder after cleaving, their entropy rises, resulting in a negative contribution.\cite{yeandel2022general} 
	The other gypsum interfaces display positive entropic contributions to the IFE, ranging from approximately 20\% to 60\%, which is much smaller than those observed for bassanite.

	\subsection{Free Energy Perturbation Methods}
	
	Free Energy Perturbation methods in molecular simulations estimate the free energy differences between two states by sampling the energy differences as the system is slightly perturbed from one state to the other.\cite{chipot2007calculating} 
	This section introduces two techniques, the test-volume and test-area methods, tailored for calculating relative IFE in fluid-solid systems.
	
	\subsubsection{Test-volume Method}
	
	The test-volume method was initially employed by Eppenga and Frenkel\cite{eppenga1984monte} to obtain the bulk pressure of isotropic and nematic phases in systems containing repulsive hard discs. 
	The method has since evolved to estimate fluid-solid relative IFE in complex molecular systems with various types of interactions,\cite{harismiadis1996efficient,vortler2000computer,de2006nature} and applied for estimating the fluid-solid relative IFE $\gamma_{\rm SF}^*$ in molecular simulations.\cite{de2006detailed,brumby2011subtleties,jimenez2012evaluation,fujiwara2014local,brumby2017structure}
	Remarkably, Fujiwara and Shibahara\cite{fujiwara2014local} extended the method for subsystems, and local pressure profiles can be obtained through their approach.
	They also introduced an instantaneous expression for local pressure components and relative IFE using the test-volume method.\cite{fujiwara2015local} This approach is particularly useful for studying time-dependent fluid-solid interfacial energies in non-equilibrium molecular dynamics simulations.

\begin{figure}[tb]
		\begin{centering}
			\includegraphics[width=0.65\textwidth]{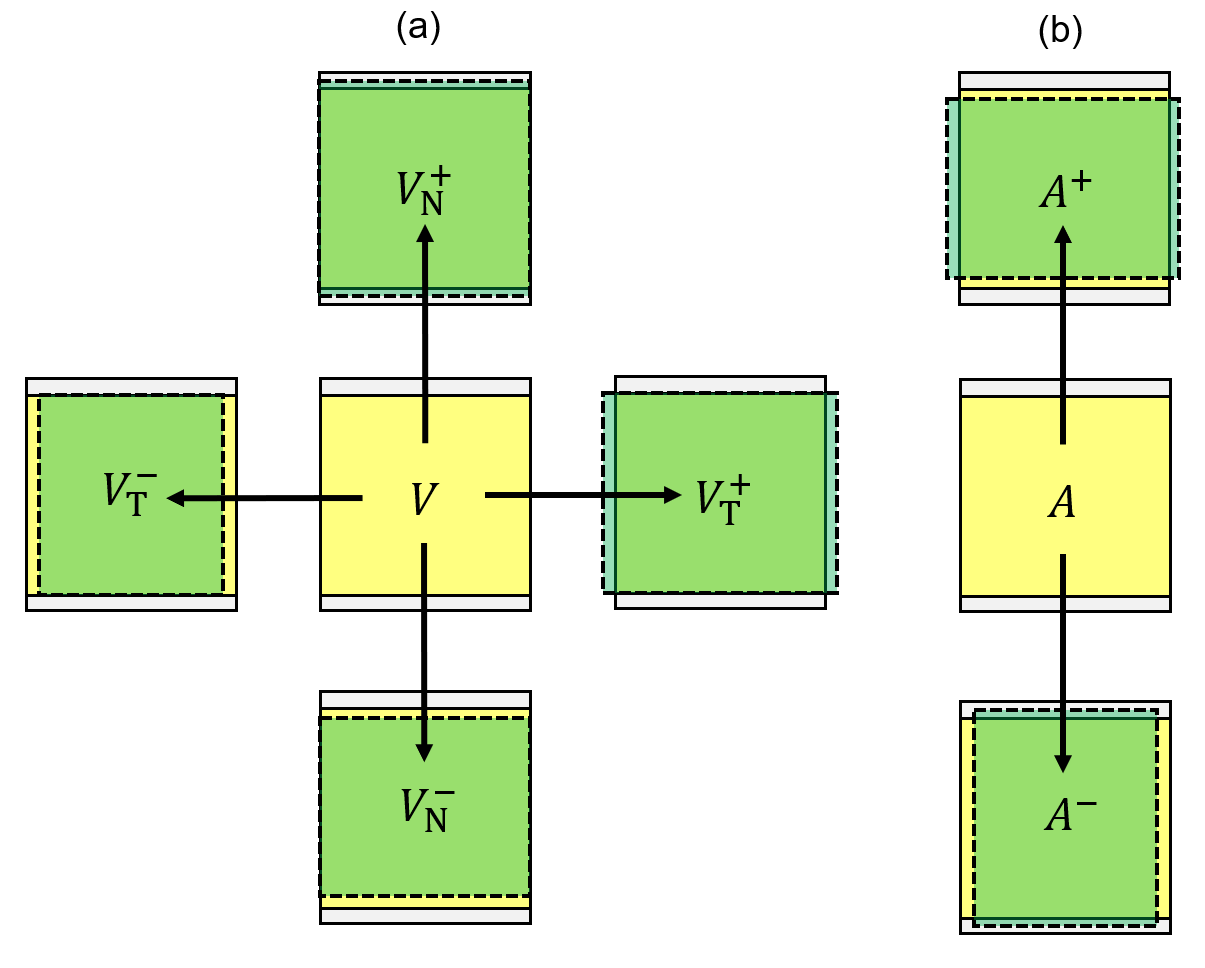}
			\caption{Sketches of free energy perturbation methods: (a) The test-volume method and  (b) the test-area method.
				Yellow regions are the fluids without any perturbations (with volume $V$ and interfacial area $A$), and the green regions within dashed boxes denote the fluids after perturbations.
				$V_{\rm N}^{+,-}$ denotes the volume of fluids under perturbations of $V$ in the normal direction with constant $A$.
				$V_{\rm T}^{+,-}$ denotes the volume of fluids under perturbations of $V$ in the tangential directions with constant $V/A$ (i.e fluid size in the normal direction).
				$A^{+,-}$ denotes the interfacial area under perturbations of $A$ with constact $V$.
				The grey regions are solids, which are treated as external fields.
			}
			\label{fig:z_f13}
		\end{centering}
	\end{figure}

	The test-volume method estimates the principle components of the pressure tensor of the fluid directly by slightly perturbing the fluid volume in directions normal and tangential to the interface. A sketch of the method for the fluid-solid system is shown in Fig. \ref{fig:z_f13}a. 
	A rigorous derivation of working equations has been given for both canonical and grand canonical ensembles based on statistical mechanics by Fujiwara and Shibahara.\cite{fujiwara2014local}
	While full derivation details are given in Ref. \citenum{fujiwara2014local}, a brief outline is presented here to give the working equation in the canonical ensemble as an example for calculating principle components of the pressure tensor:\cite{fujiwara2014local}
	\begin{equation}
		\label{eq:Pxx}
		P_{\xi\xi} = -\bigg(\cfrac{\partial F}{\partial V}\bigg)_{L_{\neq \xi} NT} = \cfrac{1}{\beta \Delta L_{\xi} A} {\rm ln} \left\langle  \bigg(1+\cfrac{\Delta V}{V} \bigg)^N \times {\rm exp}(-\beta \Delta (U+\Phi)) \right\rangle,
	\end{equation}
	where $\beta = 1/(k_{\rm B}T)$ with the Boltzmann constant $k_{\rm B}$, $\left\langle \cdots \right\rangle$ denotes the ensemble average in unperturbed system, and $\Delta$ indicates the change of certain property after perturbation. 
	$N$, $V$, $T$, $L_{\xi}$, and $A$ are fluid particle number, the volume of the fluid, temperature, size of the fluid in $\xi$-direction, and interfacial area, respectively.
	$F$, $U$, and $\Phi$ are the Helmholtz free energy, the potential energy between fluid particles, and the potential energy between fluid particles and the solid (treated as an external field), separately. 
	
	The volume perturbation changes the distance between particles in the system leading to variations in $U$ and $\Phi$. The equations for transformations of the distances between fluid particles and between fluid particles and solid atoms can be found elsewhere.\cite{gloor2005test,fujiwara2014local}
	It is also essential to note that the fluid volume is determined by the region occupied by its particles. 
	To ensure accuracy, the magnitude of volume perturbation must be optimized through convergence testing.\cite{fujiwara2014local}
	The method usually combines the central finite difference method to enhance the accuracy (see Fig. \ref{fig:z_f13}a).\cite{gloor2005test,fujiwara2014local} 
	
	After getting the principle components of the stress tensor of the inhomogeneous fluid, the fluid-solid  relative IFE can then be calculated after applying the sum rule to Bakker's equation (Eq. \ref{eq:bakker}):\cite{fujiwara2014local}
	\begin{equation}
		\gamma_{\rm SF}^* = L_{\rm z} \Big [P_{\rm zz}-\frac{1}{2}(P_{\rm xx}+P_{\rm yy})\Big ] ,
		\label{eq:bakker_sum}
	\end{equation}
	where $L_z$ is the size of the fluid in the z-direction ($i.e.,$ direction normal to the interface). 
	
\begin{figure}[tb]
		\begin{centering}
			\includegraphics[width=0.9\textwidth]{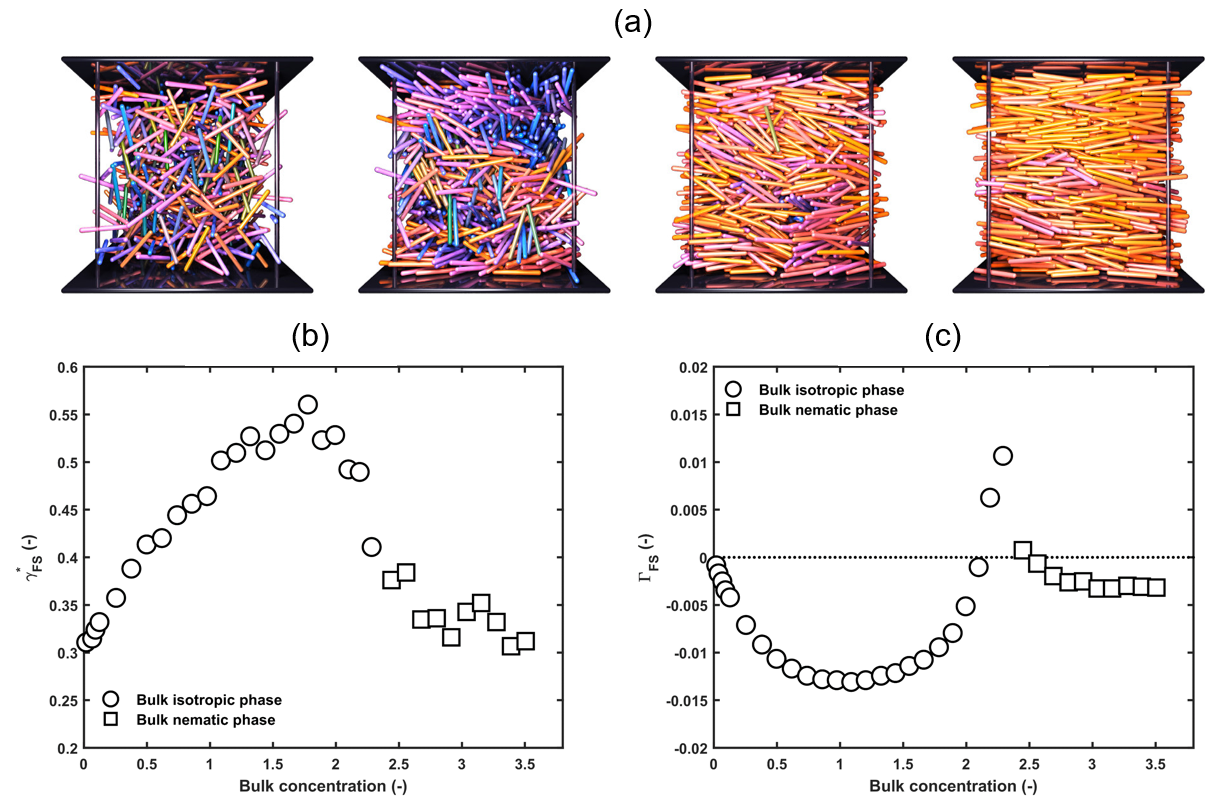}
			\caption{(a) Snapshots of hard spherocylinders within a slit pore at varying bulk concentrations: 0.7459, 2.1950, 2.4480, and 3.2792. The lower concentrations (0.7459 and 2.1950) are associated with bulk isotropic states, while the higher concentrations (2.4480 and 3.2792) correspond to bulk nematic states. 
				Colors show the different relative orientations of the particle.
				Effects of bulk concentration on (b) fluid-wall relative IFE and (c) surface adsorption.
				Figure (a) is adapted from Ref.\citenum{brumby2017structure}. The data for figures (b-c) are taken from Ref.\citenum{brumby2017structure}.}
			\label{fig:z_f14}
		\end{centering}
	\end{figure}

	The test-volume method has been recognized for its effectiveness in determining the solid-fluid interfacial tension, especially in scenarios involving high-density fluid that interacts through discontinuous potentials.
	%
	Brumby et al.\cite{brumby2017structure} applied the test-volume method to understand the relative IFE of a hard-rod fluid made of hard-spherocylinder particles in planar confinement. Such a study could offer insights into how confinement affects a dense nematic liquid-crystalline fluid, which would be valuable for applications such as liquid-crystal displays.\cite{rasing2013surfaces}
	Part of their results are summarized in Fig. \ref{fig:z_f14}. 
	Fig. \ref{fig:z_f14}a presents four representative snapshots of configurations at different bulk concentrations. As the bulk concentration increases, the bulk phase transitions from an isotropic phase to a nematic phase. 
	At small bulk concentrations, dewetting of particles at the interface of the wall occurs due to the depletion of spherocylinder particles from the impermeable wall. With an increase in bulk density, a notable change is observed in the structure of the interface.\cite{rasing2013surfaces}

	Fig. \ref{fig:z_f14}b presents the relative IFE values. 
	The trend of the relative IFEs as a function of concentration in the isotropic bulk phase exhibits a non-monotonic behavior. The maximum in relative IFEs is observed at a bulk concentration of around 2, indicating the initiation of the wetting transition and surface biaxial order.\cite{brumby2017structure}  As the density progresses further into the nematic region, the relative IFE gradually decreases and stabilizes at a constant value at high concentrations near the bulk nematic-smectic transition, estimated to occur at a bulk concentration of approximately 5.\cite{brumby2017structure}
	The comparison of relative IFEs obtained using the test-volume method was conducted against values at low bulk densities established by Mao et al.,\cite{mao1997density} where the relative IFEs are obtained from integrating the Gibbs adsorption equation with data from grand canonical simulations.
	However, employing Monte Carlo insertions in grand canonical simulations becomes increasingly challenging as density increases. It is noteworthy that the relative IFEs obtained from the test-volume method are qualitatively consistent with those from cDFT.\cite{brumby2017structure}

	The surface adsorption can be calculated by subtracting the bulk contribution from the density distributions. The relationship between surface adsorption and bulk concentrations is illustrated in Fig. \ref{fig:z_f14}c. Remarkably, the decrease in relative IFEs in bulk isotropic states correlates with a sharp increase in surface adsorption.\cite{brumby2017structure} 
	It is important to note that for the fluid-fluid interface, the non-monotonic behavior of IFE with varying density (or pressure) has been linked to the reversal of the sign of relative surface adsorption, according to the Gibbs adsorption equation.\cite{miqueu2011simultaneous,staubach2022interfacial,yang2023molecularH2}

	\subsubsection{Test-area Method}
	
	The test-area method, developed by Gloor et al.,\cite{gloor2005test} was originally applied to determine IFE in planar fluid-fluid interfaces within molecular simulations conducted in the canonical ensemble.
	The method was extended to address the relative solid-fluid IFE of inhomogeneous fluids inside slit-like pores utilizing the grand canonical\cite{miguez2012interfacial,fujiwara2014local} and canonical\cite{fujiwara2014local} ensembles based on statistical mechanics.
	The technique was also adapted to address fluids confined within cylindrical pores.\cite{blas2013extension}
	Note that Ghoufi and Malfreyt introduced a different approach to the test-area method by directly deriving the partition function. This method, known as TA2, avoids challenges related to exponential averages and the division of the surface into local elements along the normal direction.\cite{ghoufi2008expressions,ghoufi2012calculation}
	Fujiwara and Shibahara\cite{fujiwara2014local} also extended the test-area method to investigate subsystems, enabling the extraction of local pressure profiles through this refined approach. Excellent agreement of local profiles of the fluid-solid relative IFE has been reported between predictions from the perturbation method and method based on Bakker's equation using the Irving-Kirkwood contour.\cite{fujiwara2014local}

	The test-area method calculates the variation in free energy resulting from an infinitesimally small alteration in the interfacial area while maintaining the volume of the fluid. A sketch of the method for the slit pore system is shown in Fig. \ref{fig:z_f13}b. Under area perturbations in the canonical ensemble, the fluid-solid relative IFE can be calculated using the  following equation:\cite{fujiwara2014local}
	\begin{equation}
		\label{eq:TA}
		\gamma_{\rm SF}^* = \bigg(\cfrac{\partial F}{\partial A}\bigg)_{NVT} = -\cfrac{1}{\beta \Delta A} {\rm ln} \left\langle  {\rm exp}(-\beta \Delta (U+\Phi)) \right\rangle.
	\end{equation}
	
	Similar to the test-volume method, implementation of the area perturbation involves transformations of the distance between particles in the system.\cite{gloor2005test,fujiwara2014local}
	The suitable area perturbation for the method can be ascertained through a convergence test,\cite{fujiwara2014local} and the method typically integrates the central finite difference method,\cite{gloor2005test,fujiwara2014local} as illustrated in Fig. \ref{fig:z_f13}b​​.
	
	It is important to note that the formulation of the test-volume and test-area methods described above treats the solids as external fields. 
	In other words, the positions of solid particles remain unchanged during the perturbation.
	Several authors have explicitly considered the solid phase during area perturbation for determining the surface stress.\cite{nair2012molecular,d2017test} 
	The applied approach is analogous to the conventional test-area method used for fluid-fluid interfaces,\cite{gloor2005test} and it involves perturbing the positions of both fluid and solid particles.
	However, Wu and Firoozabadi pointed out that this method overlooks the deformation work of the bulk solid, potentially resulting in unphysical negative values of surface stress.\cite{wu2021calculation} They proposed a correction term that compensates for the work from the solid deformation to address the issue:
	\begin{equation}
		\label{eq:TA_WF}
		s_{\rm SF} = \gamma_{\rm TA} - \cfrac{L_{\rm z,s}}{2}\cdot\Delta\sigma,
	\end{equation}
	where $s_{\rm SF}$ denotes the surface stress (referred to as surface tension in ``Supporting Information" of Ref. \citenum{wu2021calculation}), $\gamma_{\rm TA}$ is the interfacial energy obtained from the conventional test-area method with solid particles perturbed, and the last term is the correction from solid deformation work. Factor 2 accounts for two interfaces. $L_{\rm z,s}$ is the thickness of the solid slab, and $\Delta \sigma =  (\sigma_{\rm xx}+\sigma_{\rm yy})/2-\sigma_{\rm zz}$ is the deviatoric stress of the bulk solid.
	Note that $s_{\rm SF}$ and $L_{\rm z,s}$ are determined by fitting Eq. \ref{eq:TA_WF} with various $\gamma_{\rm TA}$ from conventional test-area method and $\Delta \sigma$ determined from separate simulations of bulk solid under various stress conditions.
	With this approach, the contributions from variations in surface area and solid bulk deformations to the free energy change during test-area perturbation can be estimated and analyzed separately.\cite{wu2022reply}
	
\begin{figure}[tb]
		\begin{centering}
			\includegraphics[width=0.8\textwidth]{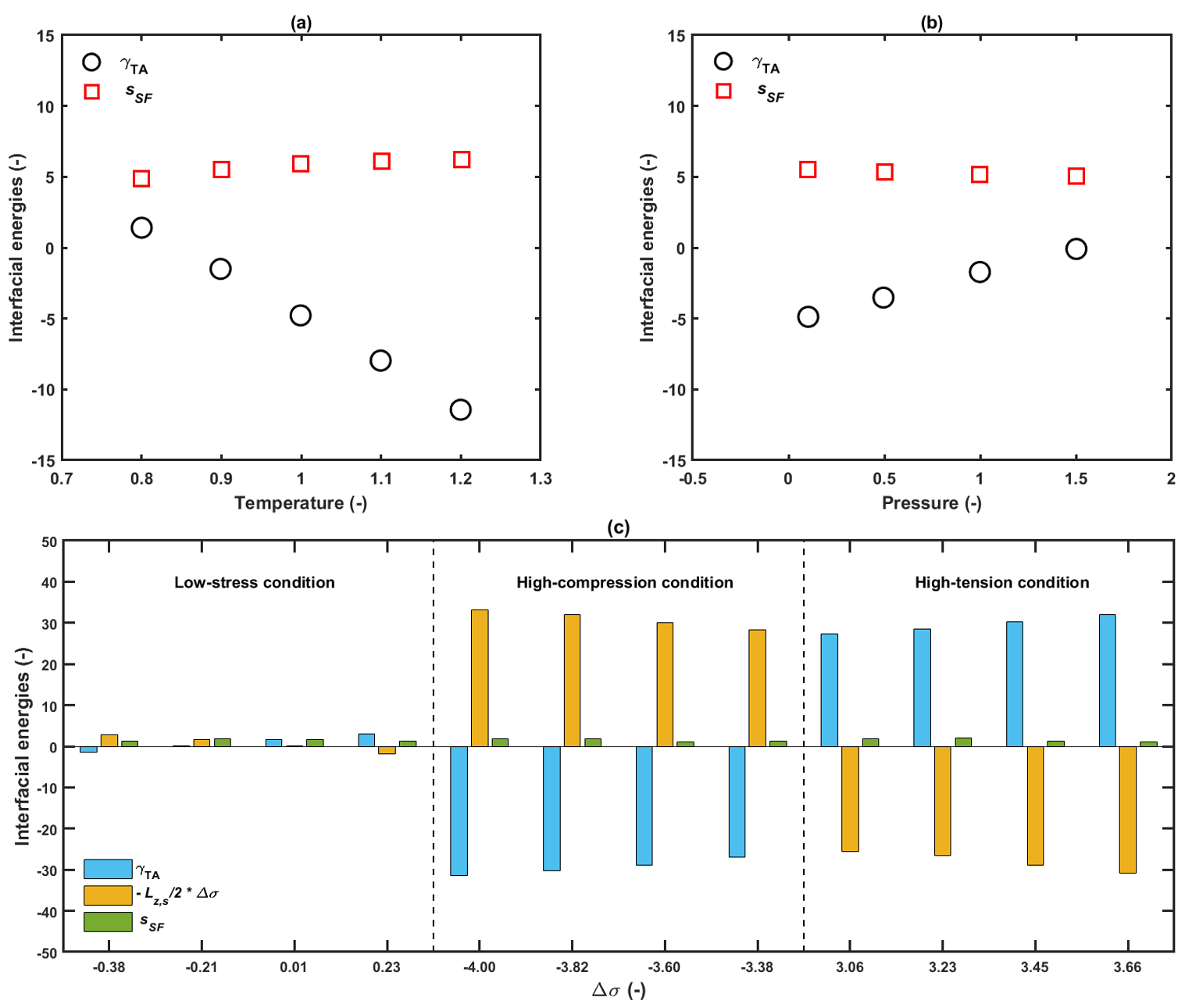}
			\caption{Interfacial energies as functions of temperature (a) and pressure (b) from the conventional test-area method and the one considers the solid deformation work. (c) Contributions of $\gamma_{\rm TA}$ obtained from the conventional test-area method and the deformation work from bulk solid $-L_{\rm z,s}/2\cdot\Delta\sigma$ to the surface stress $s_{\rm SF}$ under various deviatoric stress $\Delta \sigma$ conditions. The data are taken from Ref.\citenum{wu2021calculation}.}
			\label{fig:z_f15}
		\end{centering}
	\end{figure}

	Figs. \ref{fig:z_f15}a and b show that significant differences exist between the conventional test-area method and the one that considers solid deformation work.
	The interfacial energy $\gamma_{\rm TA}$ from the conventional test-are method ranges from negative to positive.
	The negative values violate thermodynamic stability.\cite{wu2021calculation}
	The corrected surface stresses $s_{\rm SF}$ consistently yield positive values, contrasting with the conventional method’s tendency to yield both negative and positive values.
	Remarkably, the pressure and temperature effects are opposite for those two methods.\cite{wu2021calculation} Fig. \ref{fig:z_f15}c displays the contributions of each term in Eq. \ref{eq:TA_WF} under various deviatoric stresses.
	It is observed that with higher $\Delta\sigma$, $\gamma_{\rm TA}$ deviates more from $s_{\rm SF}$.
	Under high-compression conditions, $\gamma_{\rm TA}$ values are found to be negative, and the contribution from solid deformation is positive. Opposite signs are observed under high-tension condtions.\cite{wu2021calculation}
	Very recently, Ghoufi proposed a semi-empirical method for calculating the excess free energy per unit interfacial area:\cite{ghoufi2024surface}
	\begin{equation}
		\gamma_{F}=\cfrac{F^*}{A} = \gamma_{U}-T\cdot \gamma_{S},
	\end{equation}
	where $F^*$ is the excess helmholz free energy, $T$ is the temperature, $\gamma_{U}$ is the excess internal energy per unit interfacial area ($\gamma_{U}=U^*/A$), and $\gamma_{S}$ is the excess entropy per unit interfacial area ($\gamma_{S}=S^*/A$). 
	$\gamma_{U}$ is readily available from molecular simulations. However, it is challenging to estimate $\gamma_{S}$. Ghoufi derived an expression that relates $\gamma_{S}$ with $\gamma_{U}$:\cite{ghoufi2024surface}
	\begin{equation}
		\label{eq:ghoufi}
		\big(\cfrac{\partial \gamma_{U}}{\partial T}\big)_{A,N_i^*,\epsilon_{ij}^{\Vert },\sigma_{ij}^{\perp}}=T\big(\cfrac{\partial \gamma_{S}}{\partial T}\big)_{A,N_i^*,\epsilon_{ij}^{\Vert },\sigma_{ij}^{\perp}},
	\end{equation}
	where $N_i^*$ is the excess number of molecules of component $i$, $\epsilon_{ij}^{\Vert}$ is the parallel strain, and $\sigma_{ij}^{\perp}$ is the normal stress.
	Eq. \ref{eq:ghoufi} allows to calculate $\gamma_{S}$ numerically with data of $\gamma_{U}$ under various $T$. 
	Empirical fitting the $\gamma_{U}$ with a second-order polynomial without a linear term is proposed to solve for $\gamma_{S}$.\cite{ghoufi2024surface}
	Moreover, Ghoufi rediscovered the Shuttleworth equation\cite{shuttleworth1950surface} and Dong's relation,\cite{dong2021thermodynamics} specifically applied to $\gamma_{F}$, from the first law of thermodynamics.\cite{ghoufi2024surface}
	
	Note that the method of Ghoufi does not calculate IFE $\gamma$ directly. However, $\gamma_{F}$ is related to the IFE $\gamma$ through the following relation:\cite{ghoufi2024surface}
	\begin{equation}
		\gamma=\gamma_{F}-\sum_{i}^{}\mu_i \Gamma_i^*,
	\end{equation}
	where $\Gamma_i^*=N_i^*/A$ is the surface excess of species $i$ which can be estimated based on density distributions, and $\mu_i$ is the chemical potential of component $i$. Therefore, $\gamma$ can be accessed with chemical potential being estimated elsewhere.\cite{ghoufi2024surface}
	
	The semi-empirical method offers valuable insights into $\gamma_{F}$ by integrating its energetic ($\gamma_{U}$) and entropic ($\gamma_{S}$) contributions. This approach can be applied to systems with curved interfaces without the need to calculate the local pressure tensor or introduce perturbations to the system. Additionally, the method only requires simulations at various temperatures, making it relatively more computationally efficient than other computation-intensive methods.\cite{ghoufi2024surface}
	
	The method of Ghoufi does not belong to the class of free energy perturbation methods. Nevertheless, Ghoufi compared the $s_{\rm SF}$ obtained from the semi-empirical approach with those derived from the TA2 method.\cite{ghoufi2024surface} 
	It was concluded that the test-area method is unsuitable for computing the interfacial energies of flexible solids. 
	Furthermore, there is an ongoing debate regarding the formulation of the free energy of fluid-solid system considering the solid deformation.\cite{valiya2022comment,wu2022reply,frolov2009temperature,laird2009determination} 
	Therefore, it is essential to provide a more detailed clarification of the formulation and to rigorously validate the method by conducting comparisons with other well-established techniques.
	%
	
	\section{Conclusion}
	
	This review examines methodologies for estimating the interfacial energies of fluid-solid interfaces, a critical factor for understanding wettability via Young's equation.
	An overview of experimental, semi-empirical, and first-principle theoretical methods was presented at the beginning. 
	Obtaining reliable fluid-solid IFE data using experimental and semi-empirical methods remains challenging. Furthermore, first-principle theoretical methods are limited by the intricate nature of the underlying theory and the scarcity of robust computational tools specifically designed for fluid-solid IFE calculations.
	We highlight the advantages of methods based on molecular simulations from various perspectives. 
	The primary aim of this review is to deliver an in-depth analysis of recent advancements in  molecular simulation techniques for estimating fluid-solid IFE.
	We discuss the fundamental principles, methodological developments, and practical implementations of various molecular simulation techniques, categorized into mechanical approaches—such as the contact angle method, Bakker's equation-based technique, and the Wilhelmy simulation—and thermodynamic approaches, including the cleaving wall method, the Frenkel-Ladd technique, and the test-volume/area methods.
	Notably, both the mechanical and thermodynamic approaches yield consistent values for fluid-solid IFE.\cite{yamaguchi2019interpretation,imaizumi2020wilhelmy,fujiwara2014local}
	The listed methods are suitable for calculating a range of interfacial energies of fluid-solid interfaces, including IFE, relative IFE, surface stress, and superficial tension, terms that have often been used in previous studies without clear differentiation.
	Meanwhile, selected applications of these methods are presented to gain insights into the behaviors of fluid-solid interfacial energies.
	In addition, the fluid-solid IFE is analyzed within the theoretical framework of Navascu{\'e}s and Berry\cite{navascues1977statistical} and the thought experiment based on Bakker's equation.\cite{yamaguchi2019interpretation}
	It is shown that the simulation methods based on the theory of Navascu{\'e}s and Berry\cite{navascues1977statistical} and the extended Bakker's equation derived by Nijmeijer and Leeuwen\cite{nijmeijer1990microscopic} are identical.

	%
	A key limitation in applying molecular simulation methods for fluid-solid interfacial energy estimation is the lack of widely accessible software tools and standardized benchmark examples.
	Although certain tools exist, such as the ``CLEAVING" package\cite{di2024cleaving} and the LAMMPS scripts for the Frenkel-Ladd technique\cite{yeandel2022general}), numerous other methods lack accessible, published code for replication and implementation.
	Implementing these methods may require a significant coding effort, particularly for non-expert users. 
	Therefore, developing user-friendly software packages and comprehensive benchmark examples is essential to enable wider adoption of these simulation techniques and to promote reproducibility across the research community.
	
	Another challenge arises from the computational complexity associated with calculating interfacial energies. Determining the pressure tensor is computationally demanding due to its second-order nature and slow convergence rate. 
	Methods like the contact angle and Wilhelmy simulation approaches require large-scale simulation systems and prolonged equilibrium times to account for size effects and hysteresis,\cite{jiang2019recent} significantly raising computational costs.
	Additionally, thermodynamic integration methods typically involve multiple intricate steps that require numerous simulations at various states. 
	Overcoming these computational challenges necessitates advanced numerical strategies, optimized sampling algorithms, and careful design of thermodynamic paths.
	
	There are ongoing debates rooted in certain methodologies. For example, the historical controversies surrounding the non-unique definition of the microscopic pressure tensor in Bakker's method underscore the need for clarity and consensus.\cite{shi2023perspective} 
	Further clarification is required in the formulation of test-area methods when applied to flexible solids to improve both their conceptual understanding and practical applicability.\cite{ghoufi2024surface,valiya2022comment,wu2022reply}
	Moreover, while the available methods primarily focus on classical force fields, this limitation underscores the critical need for the development of methods based on quantum mechanics.\cite{gim2018multiscale,gim2019structure} 
	Advancements in quantum mechanics-based methodologies hold the potential to broaden the scope of applications and achieve more precise prediction in complex chemical systems.
	
	The combination of experiment, simulation, and theory holds great promise for advancing the knowledge of fluid-solid interfacial energies. 
	Accurate measurement of fluid-solid IFEs at the nanoscale remains challenging, primarily due to the technical limitations of current experimental methods.
	A breakthrough in measurement techniques would greatly benefit simulation- and theory-based methods. Meanwhile, cDFT is a promising tool for estimating interfacial energies. Qualitative agreement in relative IFEs between simulations and cDFT has been reported for the nematic liquid-crystalline fluid under confinement.\cite{brumby2017structure} 
	However, cDFT's application is currently restricted by a lack of accessible, dedicated software tools, a limitation even more significant than that seen in molecular simulation methods. This highlights an urgent need for software development to fully harness cDFT's potential.
	
	The fluid-solid IFE data is scarce in the literature in contrast to the fluid-fluid IFE data.\cite{muuller2014resolving,stephan2023vapor,wan2023molecular,yang2023molecularH2}
	The insights provided in this review can broaden the application of molecular simulation methods across diverse chemical systems.
	These systems include but are not limited to, spherical and cylindrical surfaces, confined systems, charged surfaces, surfaces grafted with complex functional groups, and fluids exhibiting intricate phase behaviors. 
	Interfacial energy data from molecular simulations can provide valuable reference values that improve theoretical methods\cite{zenkiewicz2007methods,evans1979nature,wu2007density} and data-driven approaches.\cite{pollice2021data,wang2023artificial} These advances will support more accurate and efficient predictions of fluid-solid IFEs in a range of engineering and scientific applications.
	
	\bigskip
	{\bf{Acknowledgments\\[1ex]}}
	The authors acknowledge the support from the National Natural Science Foundation of China under Grant No. 42203041, the Natural Science Foundation of Jiangsu Province under Grant No. BK20221132, and the Hong Kong Scholars Program (XJ2023042).
	The authors also acknowledge the support from King Abdullah University of Science and Technology (KAUST) through the grants BAS/1/1351-01 and URF/1/5028-01. 
	The authors sincerely thank the anonymous reviewers and the editor for their valuable and constructive feedback.
	
	
	\newpage
	\bibliography{New}
	
	\clearpage
	
	\clearpage

	\clearpage
	
	\clearpage
	
	\clearpage
	
	\clearpage
	
	\clearpage
	
	\clearpage
	
	\clearpage

	\clearpage
	
	\clearpage
	
	\clearpage
	
	\clearpage
	
	\clearpage

	\clearpage
	
	\newpage
	\begin{figure}[tb]
		\begin{centering}
			\centerline{\includegraphics[width=1.0\textwidth]{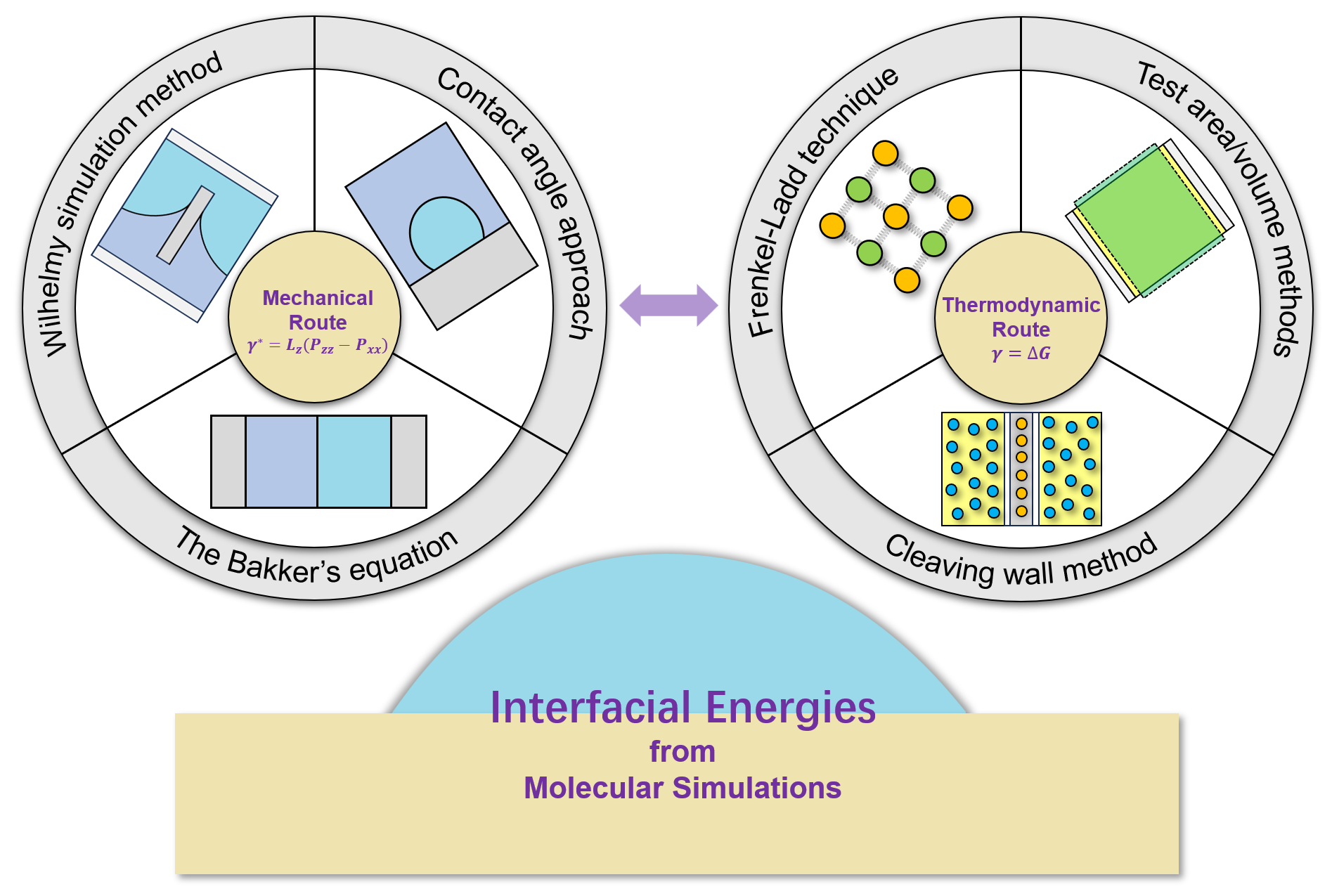}}
			\vskip0.5cm
		\end{centering}
		\vskip3cm
		{\large \bf TOC Graphic\\[2ex]}
	\end{figure}

\end{document}